\documentclass[aip,jap,amsmath,amssymb,preprint]{revtex4-1}

\usepackage{graphicx}
\usepackage{dcolumn}
\usepackage{bm}
\usepackage[utf8]{inputenc}
\usepackage[T1]{fontenc}
\usepackage{mathptmx}
\usepackage{etoolbox}
\usepackage{orcidlink}
\usepackage{array}
\usepackage{lipsum,color,soul,multirow,booktabs,textcomp,textgreek}
\usepackage{hyperref}
\hypersetup{
  colorlinks   = true, 
  urlcolor     = blue, 
  linkcolor    = blue, 
  citecolor    = blue  
}
\makeatletter
\def\@email#1#2{%
 \endgroup
 \patchcmd{\titleblock@produce}
  {\frontmatter@RRAPformat}
  {\frontmatter@RRAPformat{\produce@RRAP{*#1\href{mailto:#2}{#2}}}\frontmatter@RRAPformat}
  {}{}
}%
\makeatother

\begin{document}
\preprint{}

\title[]{Energy-dependent trapping of backscattered Helium during magnetron sputtering growth: A molecular dynamics study}


\author{Pascal Brault\, \orcidlink{0000-0002-8380-480X}}%
\email{pascal.brault@ms4all.eu}
\affiliation{GREMI, UMR7344, CNRS-Université d’Orléans, 14 rue d’Issoudun, F-45067 Orléans, France}
\affiliation{MS4ALL, 1 avenue du Champ de Mars, F-45000 Orléans, France}%
\author{Anne-Lise Thomann\,\orcidlink{0000-0002-8524-9694}}%
\email{anne-lise.thomann@univ-orleans.fr}
\affiliation{GREMI, UMR7344, CNRS-Université d’Orléans, 14 rue d’Issoudun, F-45067 Orléans, France}
\author{Alejandro Borroto\,\orcidlink{0000-0002-0246-6002}}%
\email{alejandro.borroto-ramirez@univ-orleans.fr}
\affiliation{GREMI, UMR7344, CNRS-Université d’Orléans, 14 rue d’Issoudun, F-45067 Orléans, France}%

\date{\today}


\begin{abstract}
Molecular dynamics (MD) simulations are used to investigate the trapping mechanism of Helium atoms (He) and the subsequent formation of He bubbles during the sputter deposition of aluminum (Al) and titanium (Ti) thin films in pure Helium. The simulations account for the kinetic energy distributions of sputtered metal atoms and backscattered gas species. Results indicate that He is efficiently incorporated into the growing films, leading to the formation of high-pressure, dense He bubbles, with a much higher He-to-metal atomic ratio in Ti ($\sim$ 60\%) than in Al ($\sim$ 12\%). More importantly, the simulations reveal a clear energy dependence of He trapping in Ti: background He atoms at thermal energies are essentially not incorporated, whereas efficient trapping starts at an incident energy of approximately 20 eV and increases up to a saturation regime near 100 eV. Simulated thermal desorption spectroscopy (TDS) further reveals distinct release mechanisms in the two investigated metals. Comparative simulations with argon (Ar) show weak trapping under the investigated conditions, even at kinetic energy or momentum values comparable to those of efficiently trapped He. The He and Ar sputtering conditions involve markedly different incident gas atom proportions, showing that gas incorporation cannot be only interpreted in terms of kinetic energy or momentum alone. These results are consistent with observations reported for low-energy He ion implantation in metal surfaces, suggesting similar insertion mechanisms. They clarify the atomistic conditions governing noble gas trapping during film growth and provide a basis for controlling He incorporation and porosity formation during magnetron sputtering
\end{abstract}
\maketitle

\section{Introduction}

The trapping of noble gases during thin film growth was reported as early as 1971 for Helium (He) insertion in gold films \cite{MATTOXDM1971}.More generally, discharge-gas trapping during sputter deposition has been attributed to energetic neutral gas atoms backscattered from the target, with the trapped amount strongly depending on the discharge voltage, gas-target mass ratio, pressure, substrate temperature, and deposited material \cite{Amorosi2001}.Several works have studied the incorporation level depending on the gas nature. For instance, in their study on carbon thin film deposition by a hybrid Arc-HiPIMS process, B. Akhavan \cite{Akhavan2021} highlighted that the highest the gas atom weight, the lowest amount of incorporated atoms. Due to its specific behavior in materials, it has been found that He incorporation inside the film induces significant modifications of the nano-micro-structure.  Recently, Helium bubble growth during sputter deposition in He or He/Ar sputtering gas mixtures has attracted renewed attention for synthesizing He-rich or nanoporous thin films \cite{Jia2007,Ibrahim2024,Arzac2025,Fernandez2025}. 

In an attempt to get better insight into the mechanisms involved in magnetron sputter deposition in He, some works tended to correlate the plasma characteristics with the incorporation level. It has been reported that adding small amounts percentages of Helium to the plasma has a very limited impact on its final incorporation. For example, the amount of Helium trapped in aluminum (Al) films remains extremely low, generally below 2\%, for a He fraction in the gas phase of 70\% or less \cite{Ibrahim2021}. Under these conditions, the sputtering plasma remains largely dominated by argon ions. Beyond 70\% Helium in the gas phase, the Helium content in the aluminum films increases significantly, reaching a maximum of 15\% for a plasma containing 95\% Helium \cite{Ibrahim2021}. In Ref. \cite{Jia2007}, the authors reported an approximately linear increase of the Helium concentration in aluminum films as the He/Ar flux ratio increased, reaching a saturation value of about 6.9\% He for a He/Ar ratio of 8.
The Helium is then predominantly concentrated within nanobubbles. Counter-intuitively, using pure Helium for aluminum film deposition drastically reduces the amount of incorporated Helium, which drops to less than 4\% or 5\% \cite{Ibrahim2021}. In this regime, Helium accumulates below the surface to form large bubbles that eventually burst, leading to He release. This generates a film with a highly porous and fiberform (fuzz) nanostructure, preventing massive gas trapping.\\
Beyond the gas composition, the final Helium incorporation strongly depends on the nature and growth characteristics of the deposited material. While deposition of aluminum in pure Helium (a crystalline material offering high surface adatom mobility) retains only 5\% Helium, similar silicon deposition conditions   allow for a high Helium incorporation reaching 45\% \cite{Ibrahim2024}. Silicon, which grows amorphously under these conditions, limits Helium diffusion and traps it very efficiently in oriented nanobubbles, thus forming a true solid/gas nanocomposite. 
 
In Ti films grown by magnetron sputtering even higher content levels have been reported \cite{Shi2005,Liu2007}.
The trapped Helium concentration reaches remarkably high atomic ratios, up to about 45 at.\% in one study \cite{Liu2007} and 56 at.\% in another \cite{Shi2005}, depending on the He/Ar ratio. Titanium acts as a remarkably efficient host matrix for Helium compared to aluminum. This incorporation process has been described as a continuous sub-threshold-energy implantation during film growth, allowing large amounts of He to be introduced while limiting implantation-induced displacement damage \cite{Shi2005,Liu2007}. The Helium is distributed uniformly, but its huge retention strongly affects the film structure: as concentrations rise, the lattice expands and Helium bubbles form.
The use of He/Ar mixtures has also been shown to modify film properties, partly through changes in the energetic bombardment of the growing surface \cite{Jia1999,Sarakinos2007,Rudolph2018,Ibrahim2021,Farhadizadeh2022}. However, despite the growing interest in He as a plasma-forming gas in magnetron sputtering, the atomistic mechanisms governing He trapping and bubble formation during film growth remain poorly understood. In particular, although energetic He atoms backscattered from the target have been identified as a major source of He incorporation \cite{Amorosi2001,Shi2005,Liu2007}, the incident-energy conditions required for their efficient trapping have not been clearly established. To this end, Molecular Dynamics (MD) simulations are a relevant tool for describing and analyzing He bubble formation during film deposition. Numerous MD studies have been dedicated to Helium implantation in materials, mainly for fusion energy applications \cite{Sefta2013,Pentecoste2016,Pentecoste2017}, but very few have addressed He bubble growth during the simultaneous deposition of the host material.

The objective of this work is therefore to investigate, by MD simulations, He trapping and bubble formation during the growth of Al and Ti films under sputtering-relevant conditions. The two metals are used to examine the strong matrix dependence of He incorporation, while simulated thermal desorption provides further insight into the stability of the structure and release mechanism of the trapped gas. More importantly, the incident He energy was varied from thermal background-gas energies to those characteristic of energetic backscattered atoms in order to determine the conditions required for efficient He trapping. Additional simulations with Ar are used to assess the roles of kinetic energy and momentum in gas incorporation, while considering the different incident gas fractions associated with the He and Ar sputtering conditions.

\section{Methods}
\subsection{Sputtering conditions for MD input parameters}
To define the particle fluxes and kinetic energies used as input parameters for the MD simulations, sputtering conditions representative of experimental He or Ar DC magnetron deposition of Al and Ti were considered. Depending on the target/gas combination, the target voltage measured during experiments ranged from 325 to 420 V. A total gas pressure of 1.5 Pa and a target-to-substrate distance of 11 cm were used. These sputtering conditions, summarized in Table \ref{tab:1}, were then used to determine the energy and relative fluxes of the sputtered metal atoms and backscattered gas species reaching the substrate as described below.\\

\begin{table*}[htbp]
\caption{Sputtering conditions used to determine the incident particle fluxes and mean kinetic energies, together with the main MD input parameters and final film characteristics for the Al-He, Ti-He, and Ti-Ar simulations.}
\label{tab:1}
\begin{ruledtabular}
\begin{tabular}{cccc}
 & Al-He & Ti-He & Ti-Ar \\
\hline
\# injected atoms & 22000 - 16000 & 22000 - 88000 & 10000 - 500 \\
\# atoms in film & 21300 - 2418 & 19607 - 11906 & 9119 - 12 \\
$N^{film}_{He(Ar)}/N^{film}_{metal}$ (\%) & 12 & 61 & 0.1 \\
$N^{film}_{He(Ar)}/N^{injected}_{He(Ar)}$ (\%) & 15 & 13 & 2.4 \\
$E_k^{Me}-E_k^{He,Ar}$ (eV), at substrate & 4.4 - 96 & 5.8 - 138 & 0.05 - 0.70 \\
Duration (ns) + 1ns for relaxation & 7.6 & 22.0 & 2.1 \\
Final thickness (nm) & 5 & 7 & 2.6 \\
$\gamma$ - $\beta$ & 0.23 - 0.18 & 0.07 - 0.27 & 0.55 - 0.0275 \\
\hline
\multicolumn{4}{c}{Corresponding experimental conditions} \\
\hline
$V_b$ (V) & 420 & 325 & 400 \\
$P_{He,Ar}$ (Pa) & 1.5 & 1.5 & 1.5 \\
$d_{T-S}$ (cm) & 11 & 11 & 11 \\
\end{tabular}
\end{ruledtabular}
\end{table*}

\subsection{Molecular Dynamics simulations}
Molecular Dynamics (MD) simulations describe the time evolution of an N-body atomic system by integrating Newton's equations of motion:

\begin{equation}
    \label{eq1}
     m_i \frac{d^{2} \vec{r}_i(t)}{d t^{2}} = \vec{f}_i(t), \quad \mbox{with} \quad
     \vec{f}_i(t) =  -\frac{\partial V(\vec{r}_1(t), \vec{r}_2(t), ...,\vec{r}_n(t))}{\partial \vec{r}_i(t)} 
\end{equation}

Where $\vec{r}_{i}(t)$ is the position of atom $i$ at time $t$, $m_{i}$ is its mass, and $\vec{f}_{i}(t)$ is the force acting on atom $i$ at time $t$. $V = V(\vec{r}_1, \vec{r}_2, ...,\vec{r}_n)$ is the interaction potential among all species. The atomic trajectories are then obtained by numerical integration over successive time steps.\\

 The forces between atoms are derived from the interaction potential $V(\vec{r}_1(t), \vec{r}_2(t), ...,\vec{r}_n(t))$. The Embedded-Atom Method (EAM) force field is used to describe metal-metal interactions \cite{Foiles1986,Daw1993,Foiles2012,Johnson1989,Zhou2004}, while He-metal and Ar-metal interactions are described by the Moliere interaction potential \cite{Moliere1947}. He-He and Ar-Ar interactions are modeled using Lennard-Jones potentials \cite{LJ1949a,LJ1949b}. More advanced He-He interaction potentials, such as the Juslin form \cite{Juslin2013} or Tang-Toennies \cite{Tang1995}, were tested, but no significant differences were observed for the bubble properties analyzed here.\\
To simulate the film growth under sputtering-relevant conditions, realistic initial parameters for the incoming species must first be defined. The incoming velocities of sputtered Al and Ti atoms at the substrate are determined from the sputtered atom energy distribution function (EDF) resulting from He (or Ar) bombardment of the target, at the corresponding bias voltage $V_b$, and calculated using the SRIM software\cite{Ziegler1985}. Then the calculated EDF is propagated through the background gas at pressure $P_{He}$ = $P_{Ar}$ = 1.5 Pa and target-to-substrate distance of $d_{T-S}$ = 11 cm, using a collision model accounting for gas-phase scattering \cite{Meyer1981,GrasMarti1983,Xie2014,Atmane2024} along the path between target and substrate.
The mean kinetic energy $E_k^{Me}$  (Me = Al or Ti) is determined from the resulting post-collision EDF.
Similarly, the EDF of backscattered He or Ar is calculated at the target using SRIM and is also propagated through the background gas. Accounting for the collisions allows the energy distribution reaching the substrate to be determined, from which the corresponding mean kinetic energy $E_k^{He, Ar}$ is obtained. These mean kinetic energies are then used to define the initial velocities of the incident species in the MD simulations. Finally, the relative fractions of metal and backscattered gas atoms incident on the substrate are determined using the SRIM calculated  He and Ar backscattering coefficients $\beta$ and sputtering yield $\gamma$:  \%Me =$\frac{\gamma}{\gamma+\beta}$  and \%He or \%Ar = $\frac{\beta}{\gamma+\beta}$, where \%Me and \%He or \%Ar represent the relative fractions of the incident metal and gas species, respectively. Table \ref{tab:1} summarizes the resulting input parameters used in the MD simulations.

The simulation box consists of a Fe substrate of  35 x 35 x 4 \AA$^3$, where a = 0.2886 nm is the Fe bcc lattice parameter. Above the surface, a 10 nm high vacuum zone is used for injecting both metal and He (Ar) atoms from a region located between 10 and 11.5 nm. In the simulation setup for He sputtering, the total number of injected metal atom is set to 22000 and the He atom number is set according to the corresponding $\gamma$ and $\beta$ reported in Table \ref{tab:1}. To mimic sputter deposition in the presence of backscattered He, 2000 deposition cycles were performed, each consisting of the successive release of 0.05\% of the total metal population and 0.05\% of the corresponding He population towards the substrate. The deposition cycle was repeated every 2000 timesteps using a timestep of 0.1 fs. This small timestep is required because of the relatively high kinetic energy of impinging atoms. The Fe substrate was kept at 300K using a Langevin thermostat with a damping time of 10 fs. (100 times the timestep, which is a good compromise to prevent excessive temperature fluctuations). The resulting deposition times were approximately 7.6 ns for Al and 22 ns for Ti. After deposition, the films were further relaxed for 1 ns before structural analysis. Radial distribution functions (RDFs) of He-He in the final films were calculated to characterize the local structure of He within the bubbles.\\
To further probe the thermal stability of trapped He, desorption was investigated by simulated annealing. Simulated annealing was performed by applying a temperature ramp from 300K to 2000K at a rate of 1 K / 4 ps, using a Langevin thermostat with a 100 fs damping time and 1 fs timestep. Because this heating rate is many orders of magnitude higher than that used in conventional thermal desorption experiments, the simulated desorption temperatures are interpreted comparatively rather than as a direct quantitative reproduction of experimental TDS. The evolution of the remaining He, Al, and Ti atoms was monitored to assess the relative thermal stability and release behavior of the trapped He.
To isolate the effect of the incident energy of backscattered He atoms, additional He-assisted Ti growth simulations were performed at different He kinetic energies $E_k^{He}$,
but for 10000 incoming Ti and 40000 He atoms, keeping thus the ratio H:Ti being 4:1 as displayed in Table \ref{tab:1}.  Other simulation parameters remain identical. Additional simulations using 10000 Ti atoms and 500 incident Ar atoms were performed to compare He and Ar trapping at different kinetic energies and momenta (see Table \ref{tab:1}).  These simulation sets correspond to markedly different incident gas fractions, namely 80 \% for He and 4.8 \% for Ar. In all simulations, backscattered He and Ar were treated as neutral atoms since they are known to be neutralized at the metal target when backscattered.

Simulations were carried out using the LAMMPS software \cite{Thompson2022,LAMMPS}. Snapshots were plotted using OVITO software \cite{Stukowski2009}. The crystalline phases of the simulated films were identified using the Polyhedral Template Matching method implemented in OVITO. The virial atomic stress of the simulated films was calculated during MD run \cite{Shen2004,Atmane2024} and visualized using OVITO.

\section{Results and discussion}
\subsection{He trapping and gas/solid composite growth}
\begin{figure}[ht]
    \centering
\includegraphics[scale=0.8]{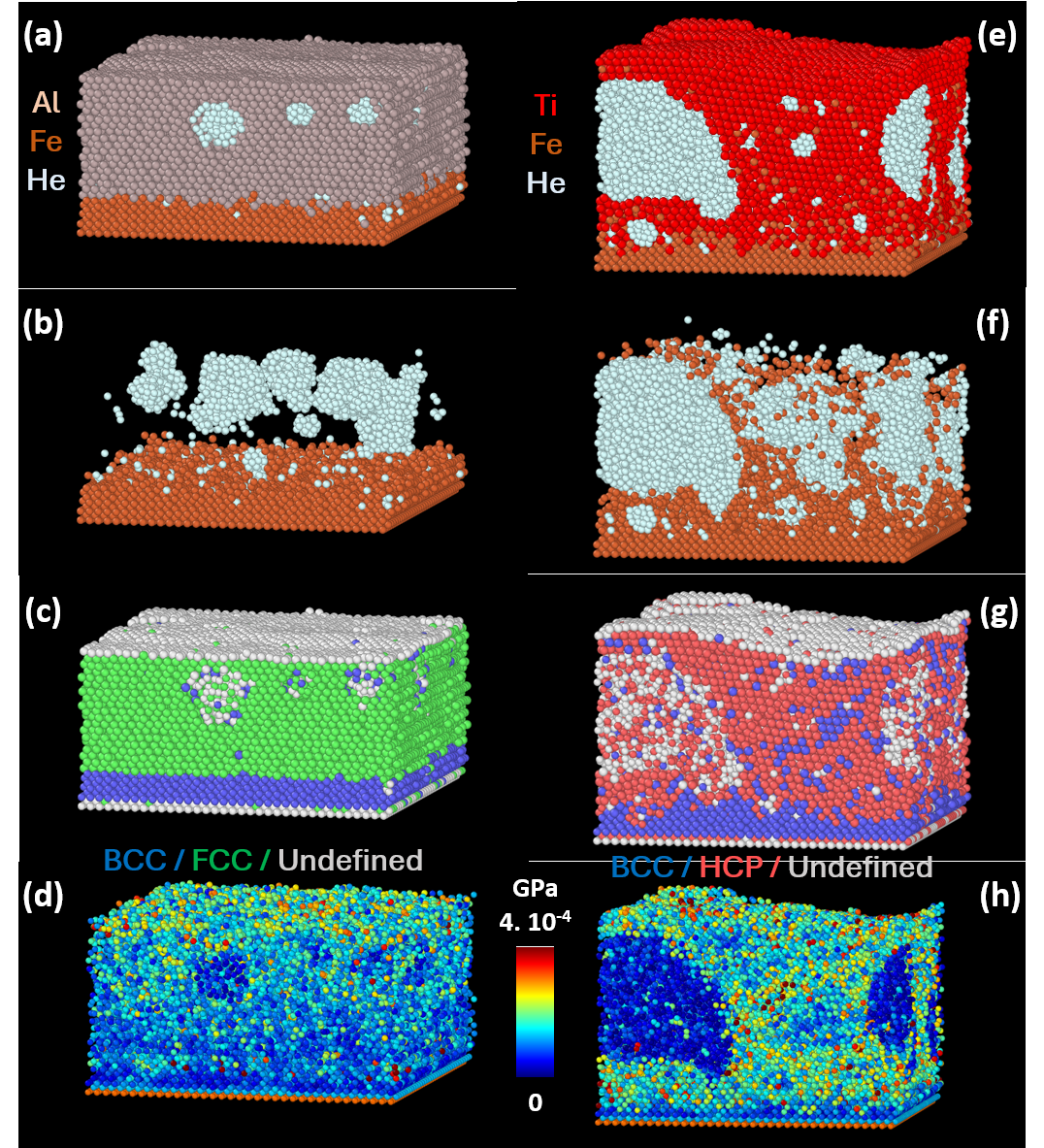}
    \caption{Final atomic configurations of the Al-He (a-d) and Ti-He (e-h) films after sputter growth and relaxation. (a,e) Complete film configurations; (b,f) configurations with Al or Ti atoms removed to highlight the distribution of trapped He; (c,g) crystalline structures of the films and substrate; and (d,h) corresponding virial atomic stress maps.}
    \label{fig:AlTi_2000}
\end{figure}

\begin{figure}[ht]
    \centering
\includegraphics[scale=0.8]{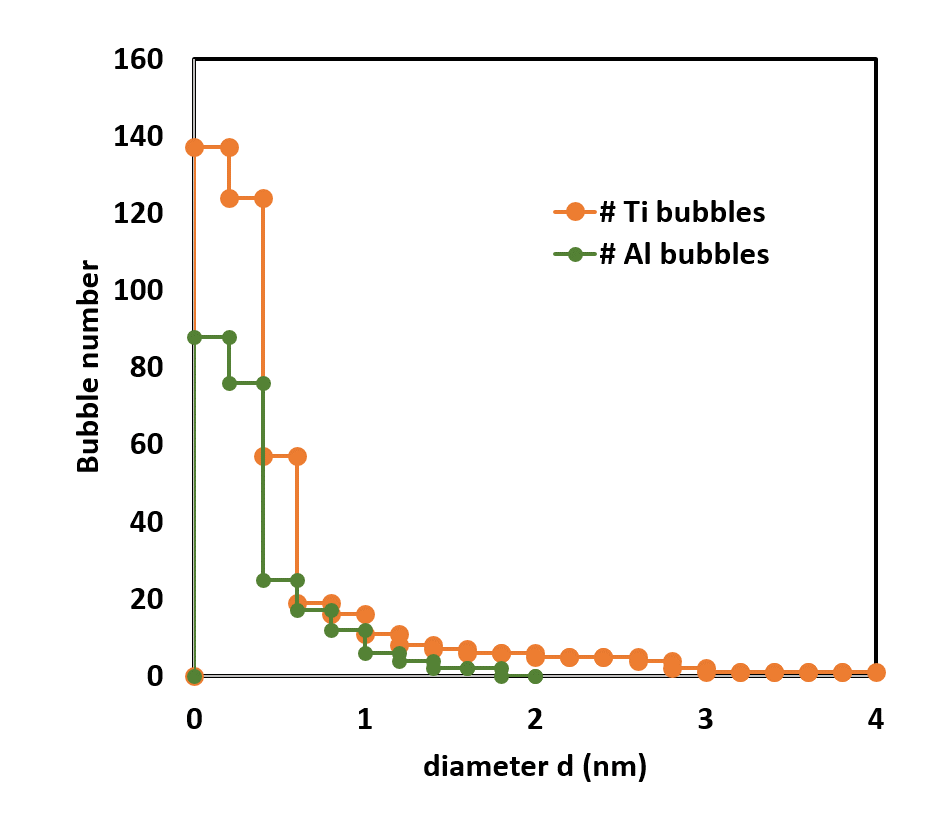}
    \caption{Size distribution of He bubbles in the Al and Ti films at the end of the deposition simulations.}
    \label{HeSize}
\end{figure}

Table \ref{tab:1} details the composition of the MD simulated films after 7.6 and 22. ns for He in Al and Ti, respectively. The He-to-metal atomic ratio in Ti is about 5 times larger than in Al, consistent with the difference in the initial He/metal ratios of the injected atoms (see Table \ref{tab:1}) directly correlated to the sputtering and backscattered rates. The amount of incorporated He in the Al film (He/Al atomic ratio of 12\% ) is close to the highest values reported experimentally \cite{Jia2007,Ibrahim2021} where He/Al ratios up to about 15\% were reached, at 95\% He in the He/Ar plasma. The He concentration in the Ti films reaches He/Ti atomic ratio of 61\%, also close to values reported in experimental works \cite{Shi2005,Liu2007} where up to 56\% He was incorporated. Although a direct quantitative comparison is limited due to the length and time scales accessible to MD, the simulations reproduce the much stronger He retention observed experimentally in Ti than in Al. \\
Figure \ref{fig:AlTi_2000} displays the final configurations for both Al and Ti deposits, shown with, Figure \ref{fig:AlTi_2000}(a,e) and without, Figure \ref{fig:AlTi_2000}(b,f), the film atoms to highlight He profile in the metal films. It also presents the atomic structure (HCP, FCC or BCC), Figures \ref{fig:AlTi_2000}(c,g), and the virial atomic stress maps in Figures \ref{fig:AlTi_2000}(d,h).  As observed in Figure \ref{fig:AlTi_2000}, the Al and Ti films are crystalline. The crystal structure of Al film is fully FCC as reported in experiments \cite{Jia2007,Ibrahim2021}. On the other hand, Ti exhibits a mixed HCP ($\alpha$-Ti) / BCC ($\beta$-Ti) metastable phase. It has been proved that the high-temperature $\beta$-Ti can be stabilized below 1155 K by the presence of foreign atoms \cite{Kolli2018,Pesode2023,Boivin2025}. The present result, suggests that the large amount of incorporated He may contribute to the stabilization of the metastable BCC fraction.\\
One should also notice that Fe substrate atoms diffuse into the growing Ti, upto the surface (Figures \ref{fig:AlTi_2000}(b,f)), whereas diffusion is limited in the case of Al growth.This may be related to the larger number and higher energy of the incident He atoms in the Ti simulation, which can enhance intermixing at the substrate-film interface.\\
Even if snapshots clearly highlight the formation of He bubbles in both metals, the sizes (as defined by the gyration diameter, since bubbles are not always spherical) and morphologies are different for Al and Ti films (Figures \ref{fig:AlTi_2000}(b,f)). Helium bubbles in Al are rather spherical with a mean diameter of about 1 nm, when only considering bubbles with more than 30 atoms (Figure \ref{HeSize}) and some He atoms and dimers mainly located at the substrate-film interface. For Ti, the He bubble mean gyration diameter (He atom number also greater than 30) is about 2 nm, although this value is strongly influenced by the very large He bubble visible in Figure \ref{fig:AlTi_2000}(f). Single He atoms and dimers are, contrary to Al, well dispersed in the film, surrounding larger bubbles. When analyzing snapshot evolution during the simulations for Al, it is seen that He bubbles grow separatetely without coalescence. On the contrary, He bubbles in Ti are quickly coalescing leading to few very large bubbles. This is consistent with the conclusions drawn from experimental results for He in Ti \cite{Shi2005,Liu2007}.\\
Stress maps in Figure \ref{fig:AlTi_2000}(d,h) reveals that He bubbles, both in Al and Ti films, present a minimum atomic stress. The stress is higher at the film surface for both films. It is also high between the bubbles and at the substrate–film interface in the case of the Ti film.\\

\begin{figure}[ht]
    \centering
\includegraphics[scale=0.8]{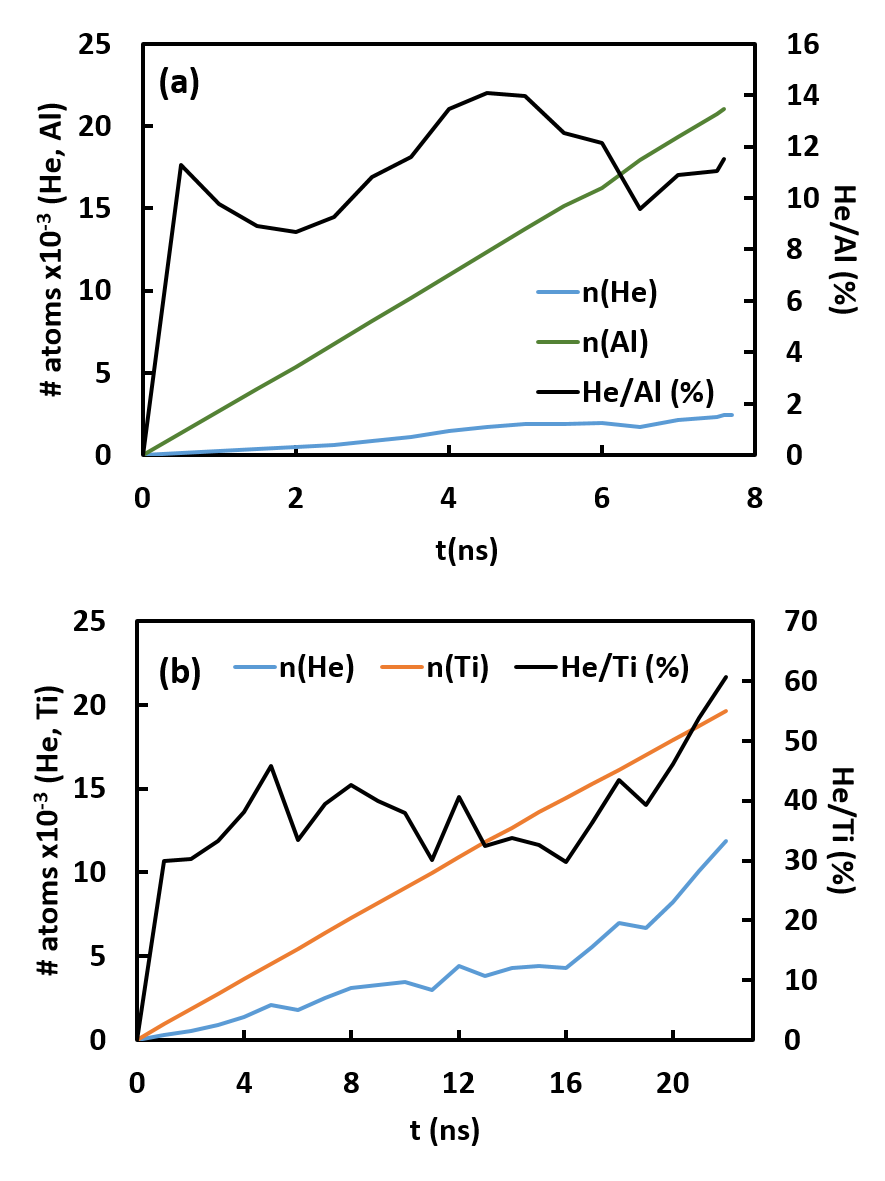}
    \caption{Time evolution of the retained metal and He atom populations during film growth: (a) Al and He atom numbers and He/Al atomic ratio; (b) Ti and He atom numbers and He/Ti atomic ratio.}
    \label{evol}
\end{figure}

In Figure \ref{evol}, the time evolution of He, Al and Ti atom counts is plotted. The number of both Al and Ti atoms evolves linearly with time, whereas the amount of retained He does not. The plots show the oscillation of the He/metal atom number during the simulation: this means that large He bubbles can break the film during growth, leading to He release at the surface. The He/Al atomic ratio appears to saturate around 12\% while in the Ti film, it initially reaches a plateau at 40\% before rising further to 60\% without any clear indication of saturation. This behavior is also encountered, both experimentally and numerically, when exposing bulk materials to He fluxes. As the He content is increased, pressurized He bubbles can reach and destabilize the surface, eventually leading to bubble bursting or surface damage \cite{Ueda2014,Pentecoste2016,Pentecoste2017}. Such surface release mechanisms have been associated with He saturation in irradiated bulk materials \cite{Pentecoste2016,Fan2023}. The present simulations suggest that a similar mechanism occurs during He-assisted growth, which supports the idea that backscattered He implantation takes place in the growing film and plays a role in He insertion.

\begin{figure}[ht]
    \centering
\includegraphics[scale=0.8]{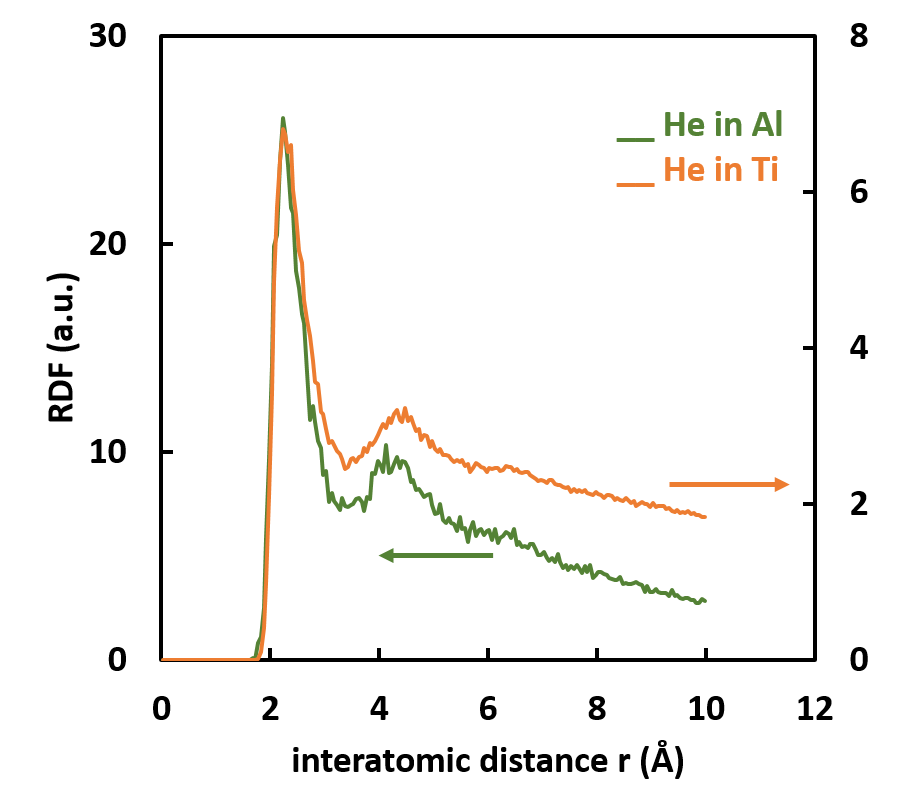}
    \caption{He-He radial distribution functions (RDFs) calculated for the Al-He and Ti-He films at the end of the deposition simulations.}
    \label{rdfHe}
\end{figure}

Radial distribution functions (RDF) of He provide information on the He bubble structure in Al and Ti.  They are displayed in Figure \ref{rdfHe}. 
The RDFs exhibit two main peaks, located at $d1_{He}$=2.25 and $d2_{He}$=4.2 \AA{} for He in Al and =4.4 \AA{} for He in Ti, respectively. A third smaller peak is emerging around $d3_{He}$= 6.5 \AA{}, indicating the possible development of some medium range order. The shape of the plots suggests that He is in a liquid-like state (two large peaks indicating broad fist and second  neighbor distances \cite{Mazo1955}), at high pressure \cite{Koci2007}.
Furthermore, assuming a mean He-He interatomic distance of $d_{He} = 2r_{He} = 0.225$ nm, the average density $\rho_{He}$ is given by $\rho_{He} = \rho_0 \left(\frac{3}{4\pi r_{He}^3}\right) \approx 167\rho_0$ nm$^{-3}$, where $\rho_0$ is the packing fraction. Using the random close-packing approximation ($\rho_0 \approx 0.64$), we obtain an effective density of $\rho_{He} \approx$ 100 atoms.nm$^{-3}$.
In their detailed MD work on He bubble formation in iron, S. M. H. Haghighat \emph{et al.} determined the He density and pressure as a function of He content and temperature \cite{Hafez2009}. At 300K, a density of about 100 atoms.nm$^{-3}$ corresponds to a pressure of about $P$ = 10 GPa. Since the He-He RDF first neighbor distances are the same in both Al and Ti, the same He density is obtained within this packing approximation, resulting in similar He bubble pressures in both films. Moreover, estimating the pressure from the RDF does not explicitly require any property of the host metal. This is also in agreement with previous MD findings where He bubble pressures, in metals, were estimated to several tens of GPa \cite{Sharma2024}, depending on bubble size.  Another way of estimating bubble pressure is the balance between the internal pressure and surface energy $\sigma$ of a spherical cavity of radius $R$ in a solid \cite{Donnelly1985,Schierholz2015}: $P =\frac{2\sigma}{R}$. Since the surface energies of Al and Ti are 1.15 and 2.0 J.m$^{-2}$ \cite{Vitos1998} with He bubble radii of 0.5 and 1 nm respectively, the estimated bubble pressures are about 4.6 GPa in Al and 4.0 GPa in Ti, of the same order as the RDF based value. Since bubble sizes and surface energies are different by nearly a factor 2, this yields a similar He bubble pressure for both Al and Ti. Both approaches therefore indicate highly pressurized He bubbles, with pressures of several GPa.

\subsection{Gas-solid composite annealing}
\begin{figure}[ht]
    \centering
\includegraphics[scale=0.9]{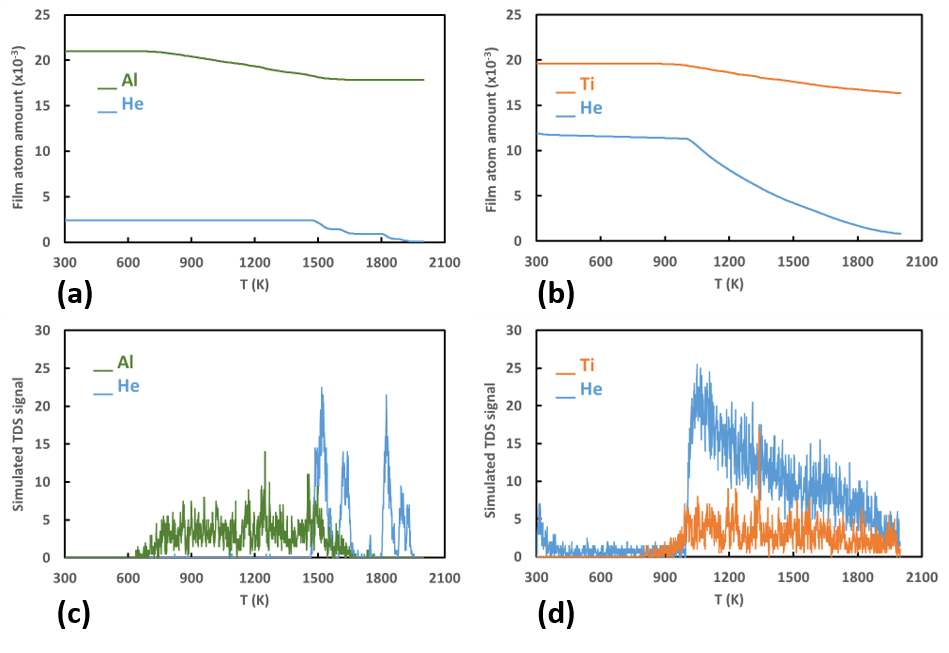}
    \caption{Evolution of the remaining film and He atoms during simulated annealing from 300 to 2000 K for (a) Al-He and (b) Ti-He films, together with the corresponding simulated TDS signals for (c) Al-He and (d) Ti-He.}
    \label{tds}
\end{figure}

\begin{figure}[ht]
    \centering
\includegraphics[scale=0.8]{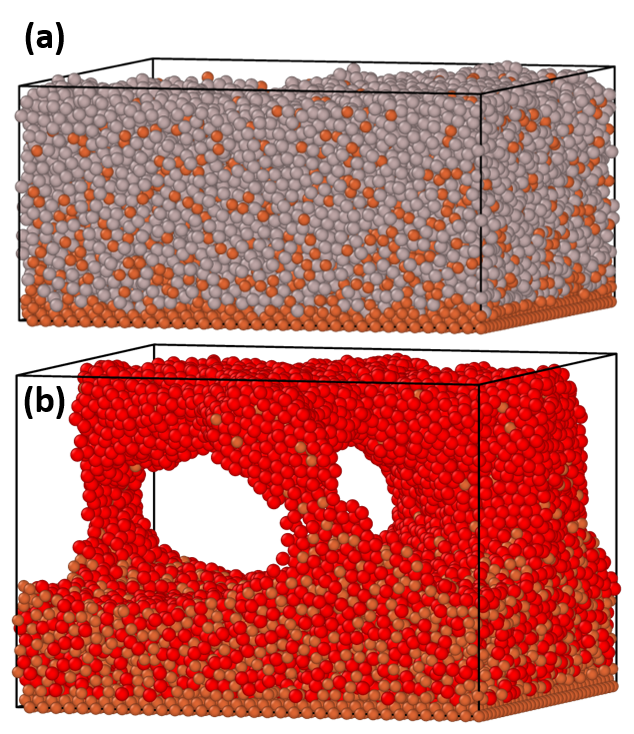}
    \caption{Final atomic configurations after simulated annealing to 2000 K for the (a) Al-He and (b) Ti-He systems.}
    \label{tds-snapshot}
\end{figure}

 Figure \ref{tds} displays the evolution of the remaining He atoms and corresponding Al and Ti film atoms, during annealing from 300 to 2000K. A deduced Thermal Desorption Spectroscopy (TDS) signal is also plotted to provide a qualitative comparison with experimental TDS data. This simulated TDS is simply obtained by calculating the first derivative of the remaining atom plots.
He desorption from the Al film requires the prior removal of top Al atoms as shown in Figure \ref{tds} (a) and (c). The fraction of removed Al is approximately 10\% from 800K (close to Al melting temperature) to 1500 K, after which no more Al desorption is recorded.  Moreover, the corresponding TDS spectrum (Figure \ref{tds} (c)) shows that He desorbs as pulses at various temperatures above 1500 K.This high He desorption temperature appears to be associated with the structural evolution of the Al film during annealing, including the preliminary removal of surface Al atoms.Unfortunately, no He desorption studies for Al coatings were found in the literature for comparison.  Regarding the Ti film, a slight and continuous removal of Ti film atoms is observed starting at 900 K (Figure \ref{tds}(b)), while He desorption begins around 950 K and proceeds continuously until nearly complete He removal is achieved (Figure \ref{tds}(d)). This desorption temperature falls within the intermediate-temperature range experimentally associated with He release from bubbles in Ti films \cite{Liu2007}, although the experimental desorption behavior strongly depends on the He content. The desorption of He starts well below the Ti melting (1940 K), confirming the role of expanding He bubbles breaking the surface. Furthermore, an examination of the final snapshots after annealing reveals that the Al filmhas evolved towards a dense He-free structure, suggesting a complete reorganization of the structure after He release, whereas the architeture of Ti film is preserved, leading to large empty voids. These behaviors can be related to the fact that He release takes place very close or far from the melting point of both metals: 930 K for Al and 1940 K for Ti, respectively.  This is in agreement with previous works showing that both the kinetic of desorption and the final state after He release greatly depend on the host material \cite{Fernandez2025}. Both films display a large amount of Fe diffusion after annealing (Figure \ref{tds-snapshot}).

\subsection{Energy condition for He gas-solid composite formation}
A question remains: during plasma sputter deposition, there are two populations of Helium, backscattered He atoms, which can retain kinetic energies above several tens of eV depending on the $P_{He}.d_{T-S}$ product, and background He gas at the process temperature ($<$ 500K, \emph{i.e.}  $E_k^{He}$ $<$ 0.05 eV ), both populations could, \emph{a priori}, contribute to Helium trapping in the growing films. Experiments show that Helium can be trapped at high He/Ti ratios up to 50\% \cite{Liu2007} while Argon is trapped at much lower level, below 0.05\% \cite{Amorosi2001}. In order to get a better insight into the incorporation process by comparing both gases, He and Ar sputter depositions at various backscattered energies were implemented. The incoming kinetic energy at the substrate position of He atoms, $E_k^{He}$,  was selected in the  range from 0.03 (background gas) to 160 eV (the value estimated for the sputtering conditions reported in Table \ref{tab:1} is 138 eV). An incoming He-to-Ti ratio of 4 was maintained (according to SRIM $\beta$ and $\gamma$ parameters). In this set of simulations, 10000 Ti atoms were released under the sputtering conditions outlined in Table \ref{tab:1} ($E_k^{Ti}$ = 5.3 eV at the Fe substrate surface). For Ar sputtering, experimental conditions and input simulation parameters are displayed in Table \ref{tab:1}.  
\begin{figure}[ht]
    \centering
\includegraphics[scale=0.7]{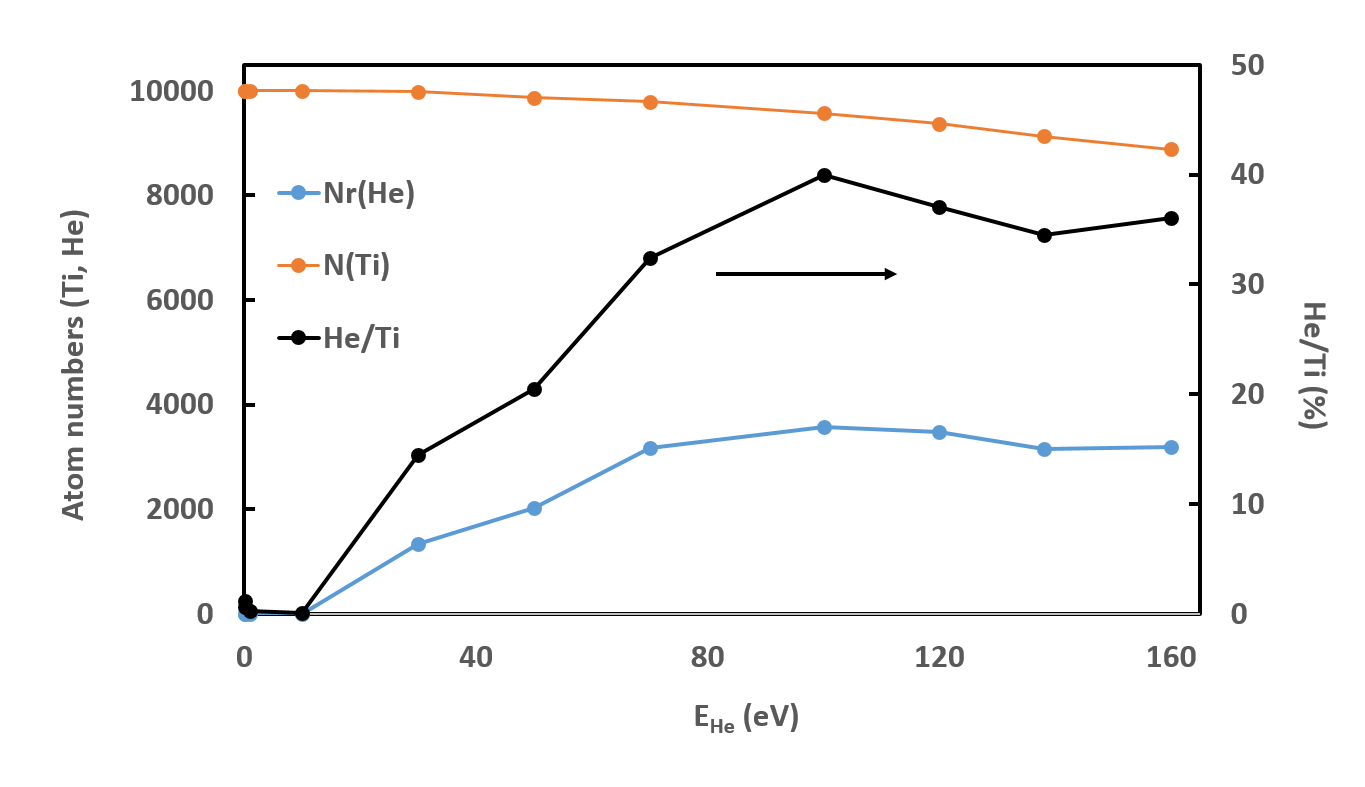}
    \caption{Dependence of He trapping on the incident He kinetic energy during Ti film growth. The retained He and Ti atom populations and the resulting He/Ti atomic ratio are shown for $E_k^{He}$=0.03-160 eV, with all other simulation parameters kept constant.}
    \label{HeEk}
\end{figure}
Figure \ref{HeEk} displays the evolution of He and Ti contents versus $E_k^{He}$. For He content, a clear onset of He trapping is observed around 20 eV. The He content increases until $E_k^{He}$ reaches 100 eV, and then saturates. This saturation occurs because He is removed when bubbles become sufficiently large to exert high pressure on the upper Ti layers. As already described above, during growth, these expanding He bubbles push toward the Ti film surface and can eventually release part of their He content. Conversely, the Ti content slightly decreases, by less than 10\% for  $E_k^{He} >$ 50 eV, due to Ti atom ejection driven by the high pressure He bubbles. Since there is almost no He trapping at 0.03 eV, the simulations indicate that the He background gas does not significantly contribute to He trapping. Efficient He incorporation therefore requires energetic incident atoms, supporting an implantation-like mechanism driven by backscattered He. This energy dependence is consistent with experimental studies on Ti showing that He retention is strongly influenced by the energy of He atoms backscattered from the target \cite{Shi2005}\\
To determine whether the energy condition identified for He is sufficient to explain the difference between He and Ar trapping, the kinetic energy of the incoming Ar atoms was first considered. Under the experimental conditions (see Table \ref{tab:1}), backscattered Ar atoms reach the substrate with a very low kinetic energy of only 0.7 eV, well below the energy threshold of 20 eV, identified for efficient He trapping. Therefore, the negligible Ar incorporation could initially be attributed to insufficient kinetic energy. To test this hypothesis, an additional simulation was performed at $E_k^{Ar}$= 100 eV, corresponding to the energy at which the trapped He content reaches saturation. Table \ref{tab:2} displays the Ar content and Ar/Ti ratios obtained at the different energies. Despite the large increase in kinetic energy, the Ar/Ti ratio remains very low, reaching only 0.11\% at 100 eV. This result indicates that increasing the Ar kinetic energy alone, under otherwise identical Ar deposition conditions, does not lead to massive gas trapping. 

\begin{table}[ht]
\caption{Retained Ar and Ti atom numbers and resulting Ar/Ti atomic ratios in the simulated Ti films as a function of the incident Ar kinetic energy.}
\label{tab:2}
\centering
\begin{tabular}{c c c c } 
 \hline \hline
  $E_{Ar}$ (eV) & \#Ti & \#Ar & Ar/Ti (\%) \\
  \hline
0.03  & 8947 & 27 & 0.30\\
0.70  & 9119 & 12 & 0.13\\
10    & 9147 & ~2 & 0.02\\
100   & 9051 & 10 & 0.11\\
\hline
\end{tabular}
\end{table}

Besides kinetic energy, other parameters must be considered when comparing He and Ar trapping. The momentum transferred by backscattered neutrals has been shown to influence the growth and phase composition of sputter-deposited films \cite{Farahani2024,Farahani2026}. In addition, the relative amount of gas incident on the growing film differs markedly between the He and Ar simulations. In the energy-dependent He simulations, 40000 He atoms were injected together with 10000 Ti atoms, whereas the Ar simulations involved only 500 Ar atoms for the same number of Ti atoms. This corresponds to injected gas fractions of 80

\begin{figure*}[ht]
    \centering
\includegraphics[scale=0.45]{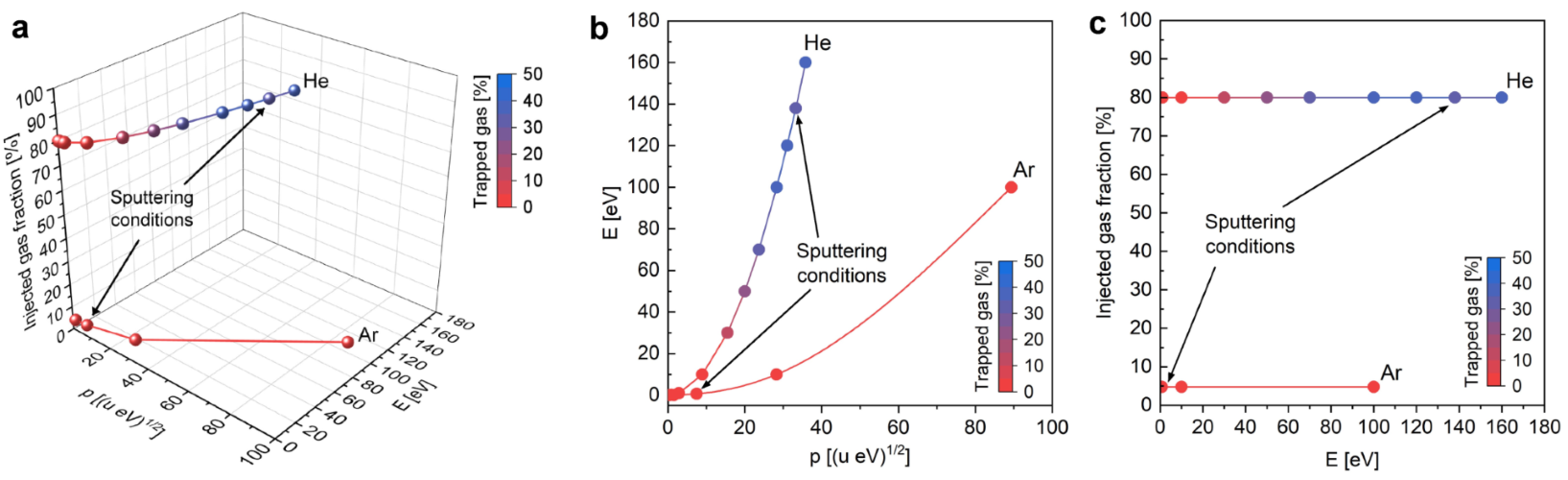}
    \caption{
Figure 8: Trapped gas-to-Ti atomic ratio as a function of (a) incident kinetic energy E, momentum $p = \sqrt{2 mE}$, and injected gas fraction; (b) kinetic energy and momentum; and (c) kinetic energy and injected gas fraction. The injected gas fraction is defined as $N_{gas}^{injected}/(N_{gas}^{injected}+N_{Ti}^{injected})$. The color scale represents the gas-to-Ti atomic ratio in the final simulated film. The conditions corresponding to the He and Ar sputter depositions are indicated by arrows.}
    \label{Ek-p}
\end{figure*}

Figure \ref{Ek-p}(a) summarizes the trapping results in the three-dimensional space defined by kinetic energy, momentum, and injected gas fraction. Its E-p projection in Figure \ref{Ek-p}(b) highlights that He atoms can reach the energy range required for efficient trapping while maintaining a relatively low momentum. In contrast, Ar at 10 eV presents a momentum comparable to that of He at 100 eV but insufficient kinetic energy for significant trapping, whereas increasing the Ar energy to 100 eV results in a much higher momentum without producing efficient incorporation. Figure \ref{Ek-p}(c) further highlights the markedly different injected gas fractions associated with the He and Ar sputtering conditions, namely 80

These differences must be considered together when comparing He and Ar incorporation. In particular, the weak Ar incorporation cannot be interpreted solely in terms of kinetic energy or momentum, since the relative amount of incident gas also differs strongly between the two sputtering conditions. Importantly, the injected gas fractions are not arbitrary simulation parameters, but directly result from the sputtering yields and backscattering coefficients associated with the corresponding sputtering conditions. Thus, while the kinetic energy was deliberately varied over a broad range, the He and Ar simulations also reflect intrinsically different incident gas fractions imposed by the sputtering process. Under the conditions investigated here, efficient trapping is observed for He, whereas Ar remains only weakly incorporated. The present results therefore indicate that gas trapping cannot be rationalized from kinetic energy or momentum alone, but results from the combined conditions under which the gas species impinge on the growing film.

\section{Conclusion}
Molecular dynamics simulations successfully modeled the trapping of He during the sputter deposition of Al and Ti films. The study demonstrates that He incorporates efficiently into these metals and forms high-pressure nanobubbles, reaching local pressures of several GPa. Ti films exhibit a significantly higher He uptake capability compared to Al. Simulated annealing and TDS analysis show distinct release mechanisms: He desorption in Ti is continuous starting at 950 K and results in a porous microstructure, whereas in Al, desorption occurs at higher temperatures as sudden pulses accompanied by significant film reorganization. These different behaviors highlight the strong influence of the host matrix on both He trapping and release mechanisms. Crucially, investigating the kinetic energy of incident He atoms revealed the onset of efficient He trapping at approximately 20 eV, with the retained He content approaching saturation near 100 eV. The negligible trapping obtained at thermal energies indicates that background He gas ($<$ 0.05 eV) does not significantly contribute to He incorporation under the investigated conditions, supporting an implantation-like mechanism driven by energetic backscattered He atoms. This is also supported by the structure similarities between sputtered films and low-kinetic energy He ion implanted surfaces. 
Comparisons with Ar sputtering show that Ar remains weakly trapped under the investigated conditions, including at kinetic energy or momentum values comparable to those of efficiently trapped He. However, the He and Ar simulations also involve markedly different incident gas fractions. The comparison therefore indicates that gas incorporation cannot be rationalized by kinetic energy or momentum alone, and that the overall conditions under which the gas species impinge on the growing film must be considered. Under the conditions investigated here, these combined effects favor efficient He trapping, whereas Ar remains only weakly incorporated. Overall, these results provide an atomistic understanding of the energetic conditions governing He trapping during sputter growth and establish a basis for controlling He incorporation in the synthesis of He-rich or nanoporous thin films by magnetron sputtering.

\subsection*{Conflict of interest}
The authors have no conflicts to disclose.
\subsection*{Author contibutions}
\textbf{Pascal Brault:} Conceptualization (equal); Investigation (lead); Formal Analysis (lead); Validation(lead); Methodology (lead);Visualization (equal);Writing/Review \& Editing (equal).\\
\textbf{Anne-Lise Thomann:} Conceptualization (equal); Investigation (supporting); Formal Analysis (equal); Validation (equal); Writing/Review \& Editing (equal); Funding Acquisition (lead).\\
\textbf{Alejandro Borroto:} Investigation (supporting); Formal Analysis (equal); Validation (equal); Visualization (equal); Writing/Review \& Editing (equal).\\
\subsection*{Data Availability Statement}
The data that support the findings of this study are available from the corresponding author upon reasonable request.\\
\bibliography{MD_HeSputt.bib}

\begin{thebibliography}{50}%
\makeatletter
\providecommand \@ifxundefined [1]{%
 \@ifx{#1\undefined}
}%
\providecommand \@ifnum [1]{%
 \ifnum #1\expandafter \@firstoftwo
 \else \expandafter \@secondoftwo
 \fi
}%
\providecommand \@ifx [1]{%
 \ifx #1\expandafter \@firstoftwo
 \else \expandafter \@secondoftwo
 \fi
}%
\providecommand \natexlab [1]{#1}%
\providecommand \enquote  [1]{``#1''}%
\providecommand \bibnamefont  [1]{#1}%
\providecommand \bibfnamefont [1]{#1}%
\providecommand \citenamefont [1]{#1}%
\providecommand \href@noop [0]{\@secondoftwo}%
\providecommand \href [0]{\begingroup \@sanitize@url \@href}%
\providecommand \@href[1]{\@@startlink{#1}\@@href}%
\providecommand \@@href[1]{\endgroup#1\@@endlink}%
\providecommand \@sanitize@url [0]{\catcode `\\12\catcode `\$12\catcode `\&12\catcode `\#12\catcode `\^12\catcode `\_12\catcode `\%12\relax}%
\providecommand \@@startlink[1]{}%
\providecommand \@@endlink[0]{}%
\providecommand \url  [0]{\begingroup\@sanitize@url \@url }%
\providecommand \@url [1]{\endgroup\@href {#1}{\urlprefix }}%
\providecommand \urlprefix  [0]{URL }%
\providecommand \Eprint [0]{\href }%
\providecommand \doibase [0]{http://dx.doi.org/}%
\providecommand \selectlanguage [0]{\@gobble}%
\providecommand \bibinfo  [0]{\@secondoftwo}%
\providecommand \bibfield  [0]{\@secondoftwo}%
\providecommand \translation [1]{[#1]}%
\providecommand \BibitemOpen [0]{}%
\providecommand \bibitemStop [0]{}%
\providecommand \bibitemNoStop [0]{.\EOS\space}%
\providecommand \EOS [0]{\spacefactor3000\relax}%
\providecommand \BibitemShut  [1]{\csname bibitem#1\endcsname}%
\let\auto@bib@innerbib\@empty
\bibitem [{\citenamefont {Mattox}\ and\ \citenamefont {Kominiak}(1971)}]{MATTOXDM1971}%
  \BibitemOpen
  \bibfield  {author} {\bibinfo {author} {\bibfnamefont {D.}~\bibnamefont {Mattox}}\ and\ \bibinfo {author} {\bibfnamefont {G.}~\bibnamefont {Kominiak}},\ }\href {\doibase 10.1116/1.1316285} {\bibfield  {journal} {\bibinfo  {journal} {Journal of Vacuum Science and Technology}\ }\textbf {\bibinfo {volume} {8}},\ \bibinfo {pages} {194} (\bibinfo {year} {1971})}\BibitemShut {NoStop}%
\bibitem [{\citenamefont {Amorosi}\ \emph {et~al.}(2001)\citenamefont {Amorosi}, \citenamefont {Anderle}, \citenamefont {Benvenuti}, \citenamefont {Calatroni}, \citenamefont {Carver}, \citenamefont {Chiggiato}, \citenamefont {Neupert},\ and\ \citenamefont {Vollenberg}}]{Amorosi2001}%
  \BibitemOpen
  \bibfield  {author} {\bibinfo {author} {\bibfnamefont {S.}~\bibnamefont {Amorosi}}, \bibinfo {author} {\bibfnamefont {M.}~\bibnamefont {Anderle}}, \bibinfo {author} {\bibfnamefont {C.}~\bibnamefont {Benvenuti}}, \bibinfo {author} {\bibfnamefont {S.}~\bibnamefont {Calatroni}}, \bibinfo {author} {\bibfnamefont {J.}~\bibnamefont {Carver}}, \bibinfo {author} {\bibfnamefont {P.}~\bibnamefont {Chiggiato}}, \bibinfo {author} {\bibfnamefont {H.}~\bibnamefont {Neupert}}, \ and\ \bibinfo {author} {\bibfnamefont {W.}~\bibnamefont {Vollenberg}},\ }\href {\doibase 10.1016/S0042-207X(00)00359-6} {\bibfield  {journal} {\bibinfo  {journal} {Vacuum}\ }\textbf {\bibinfo {volume} {60}},\ \bibinfo {pages} {89} (\bibinfo {year} {2001})}\BibitemShut {NoStop}%
\bibitem [{\citenamefont {Akhavan}\ \emph {et~al.}(2021)\citenamefont {Akhavan}, \citenamefont {Ganesan}, \citenamefont {Matthews}, \citenamefont {McKenzie},\ and\ \citenamefont {Bilek}}]{Akhavan2021}%
  \BibitemOpen
  \bibfield  {author} {\bibinfo {author} {\bibfnamefont {B.}~\bibnamefont {Akhavan}}, \bibinfo {author} {\bibfnamefont {R.}~\bibnamefont {Ganesan}}, \bibinfo {author} {\bibfnamefont {D.~T.}\ \bibnamefont {Matthews}}, \bibinfo {author} {\bibfnamefont {D.~R.}\ \bibnamefont {McKenzie}}, \ and\ \bibinfo {author} {\bibfnamefont {M.~M.}\ \bibnamefont {Bilek}},\ }\href {\doibase https://doi.org/10.1016/j.surfcoat.2021.127785} {\bibfield  {journal} {\bibinfo  {journal} {Surface and Coatings Technology}\ }\textbf {\bibinfo {volume} {427}},\ \bibinfo {pages} {127785} (\bibinfo {year} {2021})}\BibitemShut {NoStop}%
\bibitem [{\citenamefont {ping Jia}\ \emph {et~al.}(2007)\citenamefont {ping Jia}, \citenamefont {qun Shi}, \citenamefont {chun Lai},\ and\ \citenamefont {fu~Wang}}]{Jia2007}%
  \BibitemOpen
  \bibfield  {author} {\bibinfo {author} {\bibfnamefont {J.}~\bibnamefont {ping Jia}}, \bibinfo {author} {\bibfnamefont {L.}~\bibnamefont {qun Shi}}, \bibinfo {author} {\bibfnamefont {X.}~\bibnamefont {chun Lai}}, \ and\ \bibinfo {author} {\bibfnamefont {Q.}~\bibnamefont {fu~Wang}},\ }\href {\doibase 10.1016/J.NIMB.2007.06.022} {\bibfield  {journal} {\bibinfo  {journal} {Nuclear Instruments and Methods in Physics Research Section B: Beam Interactions with Materials and Atoms}\ }\textbf {\bibinfo {volume} {263}},\ \bibinfo {pages} {446} (\bibinfo {year} {2007})}\BibitemShut {NoStop}%
\bibitem [{\citenamefont {Ibrahim}\ \emph {et~al.}(2024)\citenamefont {Ibrahim}, \citenamefont {Fernández}, \citenamefont {Brault}, \citenamefont {Sauldubois}, \citenamefont {Desgardin}, \citenamefont {Caillard}, \citenamefont {Hufschmidt}, \citenamefont {de~Haro}, \citenamefont {Sauvage}, \citenamefont {Barthe},\ and\ \citenamefont {Thomann}}]{Ibrahim2024}%
  \BibitemOpen
  \bibfield  {author} {\bibinfo {author} {\bibfnamefont {S.}~\bibnamefont {Ibrahim}}, \bibinfo {author} {\bibfnamefont {A.}~\bibnamefont {Fernández}}, \bibinfo {author} {\bibfnamefont {P.}~\bibnamefont {Brault}}, \bibinfo {author} {\bibfnamefont {A.}~\bibnamefont {Sauldubois}}, \bibinfo {author} {\bibfnamefont {P.}~\bibnamefont {Desgardin}}, \bibinfo {author} {\bibfnamefont {A.}~\bibnamefont {Caillard}}, \bibinfo {author} {\bibfnamefont {D.}~\bibnamefont {Hufschmidt}}, \bibinfo {author} {\bibfnamefont {M.~C.~J.}\ \bibnamefont {de~Haro}}, \bibinfo {author} {\bibfnamefont {T.}~\bibnamefont {Sauvage}}, \bibinfo {author} {\bibfnamefont {M.~F.}\ \bibnamefont {Barthe}}, \ and\ \bibinfo {author} {\bibfnamefont {A.~L.}\ \bibnamefont {Thomann}},\ }\href {\doibase 10.1016/J.VACUUM.2024.113184} {\bibfield  {journal} {\bibinfo  {journal} {Vacuum}\ }\textbf {\bibinfo {volume} {224}},\ \bibinfo {pages} {113184} (\bibinfo {year} {2024})}\BibitemShut {NoStop}%
\bibitem [{\citenamefont {Arzac}\ \emph {et~al.}(2025)\citenamefont {Arzac}, \citenamefont {López-Viejobueno}, \citenamefont {Calvo}, \citenamefont {Ferrer}, \citenamefont {Godinho}, \citenamefont {Hufschmidt}, \citenamefont {de~Haro}, \citenamefont {Ramírez-Rico}, \citenamefont {Varela},\ and\ \citenamefont {Fernández}}]{Arzac2025}%
  \BibitemOpen
  \bibfield  {author} {\bibinfo {author} {\bibfnamefont {G.~M.}\ \bibnamefont {Arzac}}, \bibinfo {author} {\bibfnamefont {J.}~\bibnamefont {López-Viejobueno}}, \bibinfo {author} {\bibfnamefont {M.~E.}\ \bibnamefont {Calvo}}, \bibinfo {author} {\bibfnamefont {F.~J.}\ \bibnamefont {Ferrer}}, \bibinfo {author} {\bibfnamefont {V.}~\bibnamefont {Godinho}}, \bibinfo {author} {\bibfnamefont {D.}~\bibnamefont {Hufschmidt}}, \bibinfo {author} {\bibfnamefont {M.~C.~J.}\ \bibnamefont {de~Haro}}, \bibinfo {author} {\bibfnamefont {J.}~\bibnamefont {Ramírez-Rico}}, \bibinfo {author} {\bibfnamefont {F.}~\bibnamefont {Varela}}, \ and\ \bibinfo {author} {\bibfnamefont {A.}~\bibnamefont {Fernández}},\ }\href {\doibase 10.1016/j.surfcoat.2025.132850} {\bibfield  {journal} {\bibinfo  {journal} {Surface and Coatings Technology}\ }\textbf {\bibinfo {volume} {517}} (\bibinfo {year} {2025}),\ 10.1016/j.surfcoat.2025.132850}\BibitemShut {NoStop}%
\bibitem [{\citenamefont {Fernández}\ \emph {et~al.}(2025)\citenamefont {Fernández}, \citenamefont {de~Haro}, \citenamefont {Hufschmidt}, \citenamefont {Montes}, \citenamefont {Sauvage}, \citenamefont {Ferrer}, \citenamefont {Caillard}, \citenamefont {Brault},\ and\ \citenamefont {Thomann}}]{Fernandez2025}%
  \BibitemOpen
  \bibfield  {author} {\bibinfo {author} {\bibfnamefont {A.}~\bibnamefont {Fernández}}, \bibinfo {author} {\bibfnamefont {M.~C.~J.}\ \bibnamefont {de~Haro}}, \bibinfo {author} {\bibfnamefont {D.}~\bibnamefont {Hufschmidt}}, \bibinfo {author} {\bibfnamefont {O.}~\bibnamefont {Montes}}, \bibinfo {author} {\bibfnamefont {T.}~\bibnamefont {Sauvage}}, \bibinfo {author} {\bibfnamefont {F.~J.}\ \bibnamefont {Ferrer}}, \bibinfo {author} {\bibfnamefont {A.}~\bibnamefont {Caillard}}, \bibinfo {author} {\bibfnamefont {P.}~\bibnamefont {Brault}}, \ and\ \bibinfo {author} {\bibfnamefont {A.~L.}\ \bibnamefont {Thomann}},\ }\href {\doibase 10.1038/s41598-025-06889-8} {\bibfield  {journal} {\bibinfo  {journal} {Scientific Reports 2025 15:1}\ }\textbf {\bibinfo {volume} {15}},\ \bibinfo {pages} {22935} (\bibinfo {year} {2025})}\BibitemShut {NoStop}%
\bibitem [{\citenamefont {Ibrahim}\ \emph {et~al.}(2021)\citenamefont {Ibrahim}, \citenamefont {Lahboub}, \citenamefont {Brault}, \citenamefont {Petit}, \citenamefont {Caillard}, \citenamefont {Millon}, \citenamefont {Sauvage}, \citenamefont {Fernández},\ and\ \citenamefont {Thomann}}]{Ibrahim2021}%
  \BibitemOpen
  \bibfield  {author} {\bibinfo {author} {\bibfnamefont {S.}~\bibnamefont {Ibrahim}}, \bibinfo {author} {\bibfnamefont {F.~Z.}\ \bibnamefont {Lahboub}}, \bibinfo {author} {\bibfnamefont {P.}~\bibnamefont {Brault}}, \bibinfo {author} {\bibfnamefont {A.}~\bibnamefont {Petit}}, \bibinfo {author} {\bibfnamefont {A.}~\bibnamefont {Caillard}}, \bibinfo {author} {\bibfnamefont {E.}~\bibnamefont {Millon}}, \bibinfo {author} {\bibfnamefont {T.}~\bibnamefont {Sauvage}}, \bibinfo {author} {\bibfnamefont {A.}~\bibnamefont {Fernández}}, \ and\ \bibinfo {author} {\bibfnamefont {A.~L.}\ \bibnamefont {Thomann}},\ }\href {\doibase 10.1016/j.surfcoat.2021.127808} {\bibfield  {journal} {\bibinfo  {journal} {Surface and Coatings Technology}\ }\textbf {\bibinfo {volume} {426}} (\bibinfo {year} {2021}),\ 10.1016/j.surfcoat.2021.127808}\BibitemShut {NoStop}%
\bibitem [{\citenamefont {Shi}\ \emph {et~al.}(2005)\citenamefont {Shi}, \citenamefont {Liu}, \citenamefont {Xu},\ and\ \citenamefont {Zhou}}]{Shi2005}%
  \BibitemOpen
  \bibfield  {author} {\bibinfo {author} {\bibfnamefont {L.}~\bibnamefont {Shi}}, \bibinfo {author} {\bibfnamefont {C.}~\bibnamefont {Liu}}, \bibinfo {author} {\bibfnamefont {S.}~\bibnamefont {Xu}}, \ and\ \bibinfo {author} {\bibfnamefont {Z.~Y.}\ \bibnamefont {Zhou}},\ }\href {\doibase 10.1016/J.TSF.2004.11.108} {\bibfield  {journal} {\bibinfo  {journal} {Thin Solid Films}\ }\textbf {\bibinfo {volume} {479}},\ \bibinfo {pages} {52} (\bibinfo {year} {2005})}\BibitemShut {NoStop}%
\bibitem [{\citenamefont {Liu}\ \emph {et~al.}(2007)\citenamefont {Liu}, \citenamefont {Shi}, \citenamefont {Zhou}, \citenamefont {Hao}, \citenamefont {Wang}, \citenamefont {Liu},\ and\ \citenamefont {Wang}}]{Liu2007}%
  \BibitemOpen
  \bibfield  {author} {\bibinfo {author} {\bibfnamefont {C.~Z.}\ \bibnamefont {Liu}}, \bibinfo {author} {\bibfnamefont {L.~Q.}\ \bibnamefont {Shi}}, \bibinfo {author} {\bibfnamefont {Z.~Y.}\ \bibnamefont {Zhou}}, \bibinfo {author} {\bibfnamefont {X.~P.}\ \bibnamefont {Hao}}, \bibinfo {author} {\bibfnamefont {B.~Y.}\ \bibnamefont {Wang}}, \bibinfo {author} {\bibfnamefont {S.}~\bibnamefont {Liu}}, \ and\ \bibinfo {author} {\bibfnamefont {L.~B.}\ \bibnamefont {Wang}},\ }\href {\doibase 10.1088/0022-3727/40/7/044} {\bibfield  {journal} {\bibinfo  {journal} {Journal of Physics D: Applied Physics}\ }\textbf {\bibinfo {volume} {40}},\ \bibinfo {pages} {2150} (\bibinfo {year} {2007})}\BibitemShut {NoStop}%
\bibitem [{\citenamefont {Jia}, \citenamefont {Veldeman},\ and\ \citenamefont {Burgelman}(1999)}]{Jia1999}%
  \BibitemOpen
  \bibfield  {author} {\bibinfo {author} {\bibfnamefont {H.}~\bibnamefont {Jia}}, \bibinfo {author} {\bibfnamefont {J.}~\bibnamefont {Veldeman}}, \ and\ \bibinfo {author} {\bibfnamefont {M.}~\bibnamefont {Burgelman}},\ }\href {\doibase 10.1016/S0304-8853(98)00415-6} {\bibfield  {journal} {\bibinfo  {journal} {Journal of Magnetism and Magnetic Materials}\ }\textbf {\bibinfo {volume} {193}},\ \bibinfo {pages} {124} (\bibinfo {year} {1999})}\BibitemShut {NoStop}%
\bibitem [{\citenamefont {Sarakinos}\ \emph {et~al.}(2007)\citenamefont {Sarakinos}, \citenamefont {Alami}, \citenamefont {Karimi}, \citenamefont {Severin},\ and\ \citenamefont {Wuttig}}]{Sarakinos2007}%
  \BibitemOpen
  \bibfield  {author} {\bibinfo {author} {\bibfnamefont {K.}~\bibnamefont {Sarakinos}}, \bibinfo {author} {\bibfnamefont {J.}~\bibnamefont {Alami}}, \bibinfo {author} {\bibfnamefont {P.~M.}\ \bibnamefont {Karimi}}, \bibinfo {author} {\bibfnamefont {D.}~\bibnamefont {Severin}}, \ and\ \bibinfo {author} {\bibfnamefont {M.}~\bibnamefont {Wuttig}},\ }\href {\doibase 10.1088/0022-3727/40/3/014} {\bibfield  {journal} {\bibinfo  {journal} {Journal of Physics D: Applied Physics}\ }\textbf {\bibinfo {volume} {40}},\ \bibinfo {pages} {778} (\bibinfo {year} {2007})}\BibitemShut {NoStop}%
\bibitem [{\citenamefont {Rudolph}\ \emph {et~al.}(2018)\citenamefont {Rudolph}, \citenamefont {Lundin}, \citenamefont {Foy}, \citenamefont {Debongnie}, \citenamefont {Hugon},\ and\ \citenamefont {Minea}}]{Rudolph2018}%
  \BibitemOpen
  \bibfield  {author} {\bibinfo {author} {\bibfnamefont {M.}~\bibnamefont {Rudolph}}, \bibinfo {author} {\bibfnamefont {D.}~\bibnamefont {Lundin}}, \bibinfo {author} {\bibfnamefont {E.}~\bibnamefont {Foy}}, \bibinfo {author} {\bibfnamefont {M.}~\bibnamefont {Debongnie}}, \bibinfo {author} {\bibfnamefont {M.~C.}\ \bibnamefont {Hugon}}, \ and\ \bibinfo {author} {\bibfnamefont {T.}~\bibnamefont {Minea}},\ }\href {\doibase 10.1016/j.tsf.2018.05.027} {\bibfield  {journal} {\bibinfo  {journal} {Thin Solid Films}\ }\textbf {\bibinfo {volume} {658}},\ \bibinfo {pages} {46} (\bibinfo {year} {2018})}\BibitemShut {NoStop}%
\bibitem [{\citenamefont {Farhadizadeh}\ and\ \citenamefont {Kozák}(2022)}]{Farhadizadeh2022}%
  \BibitemOpen
  \bibfield  {author} {\bibinfo {author} {\bibfnamefont {A.}~\bibnamefont {Farhadizadeh}}\ and\ \bibinfo {author} {\bibfnamefont {T.}~\bibnamefont {Kozák}},\ }\href {\doibase 10.1016/j.vacuum.2021.110716} {\bibfield  {journal} {\bibinfo  {journal} {Vacuum}\ }\textbf {\bibinfo {volume} {196}} (\bibinfo {year} {2022}),\ 10.1016/j.vacuum.2021.110716}\BibitemShut {NoStop}%
\bibitem [{\citenamefont {Sefta}, \citenamefont {Juslin},\ and\ \citenamefont {Wirth}(2013)}]{Sefta2013}%
  \BibitemOpen
  \bibfield  {author} {\bibinfo {author} {\bibfnamefont {F.}~\bibnamefont {Sefta}}, \bibinfo {author} {\bibfnamefont {N.}~\bibnamefont {Juslin}}, \ and\ \bibinfo {author} {\bibfnamefont {B.~D.}\ \bibnamefont {Wirth}},\ }\href {\doibase 10.1063/1.4860315/371639} {\bibfield  {journal} {\bibinfo  {journal} {Journal of Applied Physics}\ }\textbf {\bibinfo {volume} {114}} (\bibinfo {year} {2013}),\ 10.1063/1.4860315/371639}\BibitemShut {NoStop}%
\bibitem [{\citenamefont {Pentecoste}\ \emph {et~al.}(2016)\citenamefont {Pentecoste}, \citenamefont {Brault}, \citenamefont {Thomann}, \citenamefont {Desgardin}, \citenamefont {Lecas}, \citenamefont {Belhabib}, \citenamefont {Barthe},\ and\ \citenamefont {Sauvage}}]{Pentecoste2016}%
  \BibitemOpen
  \bibfield  {author} {\bibinfo {author} {\bibfnamefont {L.}~\bibnamefont {Pentecoste}}, \bibinfo {author} {\bibfnamefont {P.}~\bibnamefont {Brault}}, \bibinfo {author} {\bibfnamefont {A.~L.}\ \bibnamefont {Thomann}}, \bibinfo {author} {\bibfnamefont {P.}~\bibnamefont {Desgardin}}, \bibinfo {author} {\bibfnamefont {T.}~\bibnamefont {Lecas}}, \bibinfo {author} {\bibfnamefont {T.}~\bibnamefont {Belhabib}}, \bibinfo {author} {\bibfnamefont {M.~F.}\ \bibnamefont {Barthe}}, \ and\ \bibinfo {author} {\bibfnamefont {T.}~\bibnamefont {Sauvage}},\ }\href {\doibase 10.1016/j.jnucmat.2015.12.017} {\bibfield  {journal} {\bibinfo  {journal} {Journal of Nuclear Materials}\ }\textbf {\bibinfo {volume} {470}},\ \bibinfo {pages} {44} (\bibinfo {year} {2016})}\BibitemShut {NoStop}%
\bibitem [{\citenamefont {Pentecoste}\ \emph {et~al.}(2017)\citenamefont {Pentecoste}, \citenamefont {Thomann}, \citenamefont {Brault}, \citenamefont {Lecas}, \citenamefont {Desgardin}, \citenamefont {Sauvage},\ and\ \citenamefont {Barthe}}]{Pentecoste2017}%
  \BibitemOpen
  \bibfield  {author} {\bibinfo {author} {\bibfnamefont {L.}~\bibnamefont {Pentecoste}}, \bibinfo {author} {\bibfnamefont {A.~L.}\ \bibnamefont {Thomann}}, \bibinfo {author} {\bibfnamefont {P.}~\bibnamefont {Brault}}, \bibinfo {author} {\bibfnamefont {T.}~\bibnamefont {Lecas}}, \bibinfo {author} {\bibfnamefont {P.}~\bibnamefont {Desgardin}}, \bibinfo {author} {\bibfnamefont {T.}~\bibnamefont {Sauvage}}, \ and\ \bibinfo {author} {\bibfnamefont {M.~F.}\ \bibnamefont {Barthe}},\ }\href {\doibase 10.1016/j.actamat.2017.08.065} {\bibfield  {journal} {\bibinfo  {journal} {Acta Materialia}\ }\textbf {\bibinfo {volume} {141}},\ \bibinfo {pages} {47} (\bibinfo {year} {2017})}\BibitemShut {NoStop}%
\bibitem [{\citenamefont {Foiles}, \citenamefont {Baskes},\ and\ \citenamefont {Daw}(1986)}]{Foiles1986}%
  \BibitemOpen
  \bibfield  {author} {\bibinfo {author} {\bibfnamefont {S.~M.}\ \bibnamefont {Foiles}}, \bibinfo {author} {\bibfnamefont {M.~I.}\ \bibnamefont {Baskes}}, \ and\ \bibinfo {author} {\bibfnamefont {M.~S.}\ \bibnamefont {Daw}},\ }\href {\doibase 10.1103/PhysRevB.33.7983} {\bibfield  {journal} {\bibinfo  {journal} {Physical Review B}\ }\textbf {\bibinfo {volume} {33}},\ \bibinfo {pages} {7983} (\bibinfo {year} {1986})}\BibitemShut {NoStop}%
\bibitem [{\citenamefont {Daw}, \citenamefont {Foiles},\ and\ \citenamefont {Baskes}(1993)}]{Daw1993}%
  \BibitemOpen
  \bibfield  {author} {\bibinfo {author} {\bibfnamefont {M.~S.}\ \bibnamefont {Daw}}, \bibinfo {author} {\bibfnamefont {S.~M.}\ \bibnamefont {Foiles}}, \ and\ \bibinfo {author} {\bibfnamefont {M.~I.}\ \bibnamefont {Baskes}},\ }\href {\doibase 10.1016/0920-2307(93)90001-U} {\bibfield  {journal} {\bibinfo  {journal} {Materials Science Reports}\ }\textbf {\bibinfo {volume} {9}},\ \bibinfo {pages} {251} (\bibinfo {year} {1993})}\BibitemShut {NoStop}%
\bibitem [{\citenamefont {Foiles}\ and\ \citenamefont {Baskes}(2012)}]{Foiles2012}%
  \BibitemOpen
  \bibfield  {author} {\bibinfo {author} {\bibfnamefont {S.~M.}\ \bibnamefont {Foiles}}\ and\ \bibinfo {author} {\bibfnamefont {M.~I.}\ \bibnamefont {Baskes}},\ }\href {\doibase 10.1557/MRS.2012.93} {\bibfield  {journal} {\bibinfo  {journal} {MRS Bulletin}\ }\textbf {\bibinfo {volume} {37}},\ \bibinfo {pages} {485} (\bibinfo {year} {2012})}\BibitemShut {NoStop}%
\bibitem [{\citenamefont {Johnson}(1989)}]{Johnson1989}%
  \BibitemOpen
  \bibfield  {author} {\bibinfo {author} {\bibfnamefont {R.~A.}\ \bibnamefont {Johnson}},\ }\href {\doibase 10.1103/PhysRevB.39.12554} {\bibfield  {journal} {\bibinfo  {journal} {Physical Review B}\ }\textbf {\bibinfo {volume} {39}},\ \bibinfo {pages} {12554} (\bibinfo {year} {1989})}\BibitemShut {NoStop}%
\bibitem [{\citenamefont {Zhou}, \citenamefont {Johnson},\ and\ \citenamefont {Wadley}(2004)}]{Zhou2004}%
  \BibitemOpen
  \bibfield  {author} {\bibinfo {author} {\bibfnamefont {X.~W.}\ \bibnamefont {Zhou}}, \bibinfo {author} {\bibfnamefont {R.~A.}\ \bibnamefont {Johnson}}, \ and\ \bibinfo {author} {\bibfnamefont {H.~N.~G.}\ \bibnamefont {Wadley}},\ }\href {\doibase 10.1103/PhysRevB.69.144113} {\bibfield  {journal} {\bibinfo  {journal} {Physical Review B}\ }\textbf {\bibinfo {volume} {69}},\ \bibinfo {pages} {144113} (\bibinfo {year} {2004})}\BibitemShut {NoStop}%
\bibitem [{\citenamefont {Moliere}(1947)}]{Moliere1947}%
  \BibitemOpen
  \bibfield  {author} {\bibinfo {author} {\bibfnamefont {G.}~\bibnamefont {Moliere}},\ }\href {\doibase 10.1515/ZNA-1947-0302/XML} {\bibfield  {journal} {\bibinfo  {journal} {Zeitschrift fur Naturforschung - Section A Journal of Physical Sciences}\ }\textbf {\bibinfo {volume} {2}},\ \bibinfo {pages} {133} (\bibinfo {year} {1947})}\BibitemShut {NoStop}%
\bibitem [{\citenamefont {Lennard-Jones}(1949{\natexlab{a}})}]{LJ1949a}%
  \BibitemOpen
  \bibfield  {author} {\bibinfo {author} {\bibfnamefont {J.~E.}\ \bibnamefont {Lennard-Jones}},\ }\href {\doibase 10.1098/RSPA.1949.0084} {\bibfield  {journal} {\bibinfo  {journal} {Proceedings of the Royal Society of London. A. Mathematical and Physical Sciences}\ }\textbf {\bibinfo {volume} {198}},\ \bibinfo {pages} {14} (\bibinfo {year} {1949}{\natexlab{a}})}\BibitemShut {NoStop}%
\bibitem [{\citenamefont {Lennard-Jones}(1949{\natexlab{b}})}]{LJ1949b}%
  \BibitemOpen
  \bibfield  {author} {\bibinfo {author} {\bibfnamefont {J.~E.}\ \bibnamefont {Lennard-Jones}},\ }\href {\doibase 10.1098/RSPA.1949.0083} {\bibfield  {journal} {\bibinfo  {journal} {Proceedings of the Royal Society of London. A. Mathematical and Physical Sciences}\ }\textbf {\bibinfo {volume} {198}},\ \bibinfo {pages} {1} (\bibinfo {year} {1949}{\natexlab{b}})}\BibitemShut {NoStop}%
\bibitem [{\citenamefont {Juslin}\ and\ \citenamefont {Wirth}(2013)}]{Juslin2013}%
  \BibitemOpen
  \bibfield  {author} {\bibinfo {author} {\bibfnamefont {N.}~\bibnamefont {Juslin}}\ and\ \bibinfo {author} {\bibfnamefont {B.~D.}\ \bibnamefont {Wirth}},\ }\href {\doibase 10.1016/J.JNUCMAT.2012.07.023} {\bibfield  {journal} {\bibinfo  {journal} {Journal of Nuclear Materials}\ }\textbf {\bibinfo {volume} {432}},\ \bibinfo {pages} {61} (\bibinfo {year} {2013})}\BibitemShut {NoStop}%
\bibitem [{\citenamefont {Tang}, \citenamefont {Toennies},\ and\ \citenamefont {Yiu}(1995)}]{Tang1995}%
  \BibitemOpen
  \bibfield  {author} {\bibinfo {author} {\bibfnamefont {K.~T.}\ \bibnamefont {Tang}}, \bibinfo {author} {\bibfnamefont {J.~P.}\ \bibnamefont {Toennies}}, \ and\ \bibinfo {author} {\bibfnamefont {C.~L.}\ \bibnamefont {Yiu}},\ }\href {\doibase 10.1103/PhysRevLett.74.1546} {\bibfield  {journal} {\bibinfo  {journal} {Physical Review Letters}\ }\textbf {\bibinfo {volume} {74}},\ \bibinfo {pages} {1546} (\bibinfo {year} {1995})}\BibitemShut {NoStop}%
\bibitem [{\citenamefont {Ziegler}\ and\ \citenamefont {Biersack}(1985)}]{Ziegler1985}%
  \BibitemOpen
  \bibfield  {author} {\bibinfo {author} {\bibfnamefont {J.~F.}\ \bibnamefont {Ziegler}}\ and\ \bibinfo {author} {\bibfnamefont {J.~P.}\ \bibnamefont {Biersack}},\ }\href {\doibase 10.1007/978-1-4615-8103-1_3} {\bibfield  {journal} {\bibinfo  {journal} {Treatise on Heavy-Ion Science}\ ,\ \bibinfo {pages} {93}} (\bibinfo {year} {1985})}\BibitemShut {NoStop}%
\bibitem [{\citenamefont {Meyer}, \citenamefont {Schuller},\ and\ \citenamefont {Falco}(1981)}]{Meyer1981}%
  \BibitemOpen
  \bibfield  {author} {\bibinfo {author} {\bibfnamefont {K.}~\bibnamefont {Meyer}}, \bibinfo {author} {\bibfnamefont {I.~K.}\ \bibnamefont {Schuller}}, \ and\ \bibinfo {author} {\bibfnamefont {C.~M.}\ \bibnamefont {Falco}},\ }\href {\doibase 10.1063/1.329473} {\bibfield  {journal} {\bibinfo  {journal} {Journal of Applied Physics}\ }\textbf {\bibinfo {volume} {52}},\ \bibinfo {pages} {5803} (\bibinfo {year} {1981})}\BibitemShut {NoStop}%
\bibitem [{\citenamefont {Gras-Marti}\ and\ \citenamefont {Valles-Abarca}(1983)}]{GrasMarti1983}%
  \BibitemOpen
  \bibfield  {author} {\bibinfo {author} {\bibfnamefont {A.}~\bibnamefont {Gras-Marti}}\ and\ \bibinfo {author} {\bibfnamefont {J.~A.}\ \bibnamefont {Valles-Abarca}},\ }\href {\doibase 10.1063/1.332113} {\bibfield  {journal} {\bibinfo  {journal} {Journal of Applied Physics}\ }\textbf {\bibinfo {volume} {54}},\ \bibinfo {pages} {1071} (\bibinfo {year} {1983})}\BibitemShut {NoStop}%
\bibitem [{\citenamefont {Xie}\ \emph {et~al.}(2014)\citenamefont {Xie}, \citenamefont {Brault}, \citenamefont {Bauchire}, \citenamefont {Thomann},\ and\ \citenamefont {Bedra}}]{Xie2014}%
  \BibitemOpen
  \bibfield  {author} {\bibinfo {author} {\bibfnamefont {L.}~\bibnamefont {Xie}}, \bibinfo {author} {\bibfnamefont {P.}~\bibnamefont {Brault}}, \bibinfo {author} {\bibfnamefont {J.~M.}\ \bibnamefont {Bauchire}}, \bibinfo {author} {\bibfnamefont {A.~L.}\ \bibnamefont {Thomann}}, \ and\ \bibinfo {author} {\bibfnamefont {L.}~\bibnamefont {Bedra}},\ }\href {\doibase 10.1088/0022-3727/47/22/224004} {\bibfield  {journal} {\bibinfo  {journal} {Journal of Physics D: Applied Physics}\ }\textbf {\bibinfo {volume} {47}},\ \bibinfo {pages} {224004} (\bibinfo {year} {2014})}\BibitemShut {NoStop}%
\bibitem [{\citenamefont {Atmane}\ \emph {et~al.}(2024)\citenamefont {Atmane}, \citenamefont {Maroussiak}, \citenamefont {Caillard}, \citenamefont {Thomann}, \citenamefont {Kateb}, \citenamefont {Gudmundsson},\ and\ \citenamefont {Brault}}]{Atmane2024}%
  \BibitemOpen
  \bibfield  {author} {\bibinfo {author} {\bibfnamefont {S.}~\bibnamefont {Atmane}}, \bibinfo {author} {\bibfnamefont {A.}~\bibnamefont {Maroussiak}}, \bibinfo {author} {\bibfnamefont {A.}~\bibnamefont {Caillard}}, \bibinfo {author} {\bibfnamefont {A.-L.}\ \bibnamefont {Thomann}}, \bibinfo {author} {\bibfnamefont {M.}~\bibnamefont {Kateb}}, \bibinfo {author} {\bibfnamefont {J.~T.}\ \bibnamefont {Gudmundsson}}, \ and\ \bibinfo {author} {\bibfnamefont {P.}~\bibnamefont {Brault}},\ }\href {\doibase 10.1116/6.0004134} {\bibfield  {journal} {\bibinfo  {journal} {Journal of Vacuum Science \& Technology A}\ }\textbf {\bibinfo {volume} {42}},\ \bibinfo {pages} {60401} (\bibinfo {year} {2024})}\BibitemShut {NoStop}%
\bibitem [{\citenamefont {Thompson}\ \emph {et~al.}(2022)\citenamefont {Thompson}, \citenamefont {Aktulga}, \citenamefont {Berger}, \citenamefont {Bolintineanu}, \citenamefont {Brown}, \citenamefont {Crozier}, \citenamefont {in~'t Veld}, \citenamefont {Kohlmeyer}, \citenamefont {Moore}, \citenamefont {Nguyen}, \citenamefont {Shan}, \citenamefont {Stevens}, \citenamefont {Tranchida}, \citenamefont {Trott},\ and\ \citenamefont {Plimpton}}]{Thompson2022}%
  \BibitemOpen
  \bibfield  {author} {\bibinfo {author} {\bibfnamefont {A.~P.}\ \bibnamefont {Thompson}}, \bibinfo {author} {\bibfnamefont {H.~M.}\ \bibnamefont {Aktulga}}, \bibinfo {author} {\bibfnamefont {R.}~\bibnamefont {Berger}}, \bibinfo {author} {\bibfnamefont {D.~S.}\ \bibnamefont {Bolintineanu}}, \bibinfo {author} {\bibfnamefont {W.~M.}\ \bibnamefont {Brown}}, \bibinfo {author} {\bibfnamefont {P.~S.}\ \bibnamefont {Crozier}}, \bibinfo {author} {\bibfnamefont {P.~J.}\ \bibnamefont {in~'t Veld}}, \bibinfo {author} {\bibfnamefont {A.}~\bibnamefont {Kohlmeyer}}, \bibinfo {author} {\bibfnamefont {S.~G.}\ \bibnamefont {Moore}}, \bibinfo {author} {\bibfnamefont {T.~D.}\ \bibnamefont {Nguyen}}, \bibinfo {author} {\bibfnamefont {R.}~\bibnamefont {Shan}}, \bibinfo {author} {\bibfnamefont {M.~J.}\ \bibnamefont {Stevens}}, \bibinfo {author} {\bibfnamefont {J.}~\bibnamefont {Tranchida}}, \bibinfo {author} {\bibfnamefont {C.}~\bibnamefont {Trott}}, \ and\ \bibinfo {author} {\bibfnamefont {S.~J.}\ \bibnamefont {Plimpton}},\ }\href {\doibase 10.1016/J.CPC.2021.108171} {\bibfield  {journal} {\bibinfo  {journal} {Computer Physics Communications}\ }\textbf {\bibinfo {volume} {271}},\ \bibinfo {pages} {108171} (\bibinfo {year} {2022})}\BibitemShut {NoStop}%
\bibitem [{LAM(2024)}]{LAMMPS}%
  \BibitemOpen
  \href@noop {} {}\bibinfo {howpublished} {\url{https://www.lammps.org }} (\bibinfo {year} {2024}),\ \bibinfo {note} {last visited on 2024/08/23}\BibitemShut {NoStop}%
\bibitem [{\citenamefont {Stukowski}(2009)}]{Stukowski2009}%
  \BibitemOpen
  \bibfield  {author} {\bibinfo {author} {\bibfnamefont {A.}~\bibnamefont {Stukowski}},\ }\href {\doibase 10.1088/0965-0393/18/1/015012} {\bibfield  {journal} {\bibinfo  {journal} {Modelling and Simulation in Materials Science and Engineering}\ }\textbf {\bibinfo {volume} {18}},\ \bibinfo {pages} {015012} (\bibinfo {year} {2009})}\BibitemShut {NoStop}%
\bibitem [{\citenamefont {Shen}\ and\ \citenamefont {Atluri}(2004)}]{Shen2004}%
  \BibitemOpen
  \bibfield  {author} {\bibinfo {author} {\bibfnamefont {S.}~\bibnamefont {Shen}}\ and\ \bibinfo {author} {\bibfnamefont {S.~N.}\ \bibnamefont {Atluri}},\ }\href@noop {} {\bibfield  {journal} {\bibinfo  {journal} {Tech Science Press cmes}\ }\textbf {\bibinfo {volume} {6}},\ \bibinfo {pages} {91} (\bibinfo {year} {2004})}\BibitemShut {NoStop}%
\bibitem [{\citenamefont {Kolli}\ and\ \citenamefont {Devaraj}(2018)}]{Kolli2018}%
  \BibitemOpen
  \bibfield  {author} {\bibinfo {author} {\bibfnamefont {R.~P.}\ \bibnamefont {Kolli}}\ and\ \bibinfo {author} {\bibfnamefont {A.}~\bibnamefont {Devaraj}},\ }\href {\doibase 10.3390/MET8070506} {\bibfield  {journal} {\bibinfo  {journal} {Metals 2018, Vol. 8, Page 506}\ }\textbf {\bibinfo {volume} {8}},\ \bibinfo {pages} {506} (\bibinfo {year} {2018})}\BibitemShut {NoStop}%
\bibitem [{\citenamefont {Pesode}\ and\ \citenamefont {Barve}(2023)}]{Pesode2023}%
  \BibitemOpen
  \bibfield  {author} {\bibinfo {author} {\bibfnamefont {P.}~\bibnamefont {Pesode}}\ and\ \bibinfo {author} {\bibfnamefont {S.}~\bibnamefont {Barve}},\ }\href {\doibase 10.1186/S44147-023-00196-7} {\bibfield  {journal} {\bibinfo  {journal} {Journal of Engineering and Applied Science 2023 70:1}\ }\textbf {\bibinfo {volume} {70}},\ \bibinfo {pages} {25} (\bibinfo {year} {2023})}\BibitemShut {NoStop}%
\bibitem [{\citenamefont {Boivin}\ \emph {et~al.}(2025)\citenamefont {Boivin}, \citenamefont {Brault}, \citenamefont {Jagodar}, \citenamefont {Vaubois}, \citenamefont {Menou}, \citenamefont {Aspe}, \citenamefont {Caillard}, \citenamefont {Andreazza}, \citenamefont {Cavarroc},\ and\ \citenamefont {Thomann}}]{Boivin2025}%
  \BibitemOpen
  \bibfield  {author} {\bibinfo {author} {\bibfnamefont {D.}~\bibnamefont {Boivin}}, \bibinfo {author} {\bibfnamefont {P.}~\bibnamefont {Brault}}, \bibinfo {author} {\bibfnamefont {A.}~\bibnamefont {Jagodar}}, \bibinfo {author} {\bibfnamefont {T.}~\bibnamefont {Vaubois}}, \bibinfo {author} {\bibfnamefont {E.}~\bibnamefont {Menou}}, \bibinfo {author} {\bibfnamefont {B.}~\bibnamefont {Aspe}}, \bibinfo {author} {\bibfnamefont {A.}~\bibnamefont {Caillard}}, \bibinfo {author} {\bibfnamefont {P.}~\bibnamefont {Andreazza}}, \bibinfo {author} {\bibfnamefont {M.}~\bibnamefont {Cavarroc}}, \ and\ \bibinfo {author} {\bibfnamefont {A.-L.}\ \bibnamefont {Thomann}},\ }\href {\doibase 10.1063/5.0224901} {\bibfield  {journal} {\bibinfo  {journal} {Journal of Applied Physics}\ }\textbf {\bibinfo {volume} {137}},\ \bibinfo {pages} {145301} (\bibinfo {year} {2025})}\BibitemShut {NoStop}%
\bibitem [{\citenamefont {Ueda}\ \emph {et~al.}(2014)\citenamefont {Ueda}, \citenamefont {Coenen}, \citenamefont {Temmerman}, \citenamefont {Doerner}, \citenamefont {Linke}, \citenamefont {Philipps},\ and\ \citenamefont {Tsitrone}}]{Ueda2014}%
  \BibitemOpen
  \bibfield  {author} {\bibinfo {author} {\bibfnamefont {Y.}~\bibnamefont {Ueda}}, \bibinfo {author} {\bibfnamefont {J.~W.}\ \bibnamefont {Coenen}}, \bibinfo {author} {\bibfnamefont {G.~D.}\ \bibnamefont {Temmerman}}, \bibinfo {author} {\bibfnamefont {R.~P.}\ \bibnamefont {Doerner}}, \bibinfo {author} {\bibfnamefont {J.}~\bibnamefont {Linke}}, \bibinfo {author} {\bibfnamefont {V.}~\bibnamefont {Philipps}}, \ and\ \bibinfo {author} {\bibfnamefont {E.}~\bibnamefont {Tsitrone}},\ }\href {\doibase 10.1016/J.FUSENGDES.2014.02.078} {\bibfield  {journal} {\bibinfo  {journal} {Fusion Engineering and Design}\ }\textbf {\bibinfo {volume} {89}},\ \bibinfo {pages} {901} (\bibinfo {year} {2014})}\BibitemShut {NoStop}%
\bibitem [{\citenamefont {Fan}\ \emph {et~al.}(2023)\citenamefont {Fan}, \citenamefont {Pan}, \citenamefont {Hu}, \citenamefont {He},\ and\ \citenamefont {Huang}}]{Fan2023}%
  \BibitemOpen
  \bibfield  {author} {\bibinfo {author} {\bibfnamefont {C.}~\bibnamefont {Fan}}, \bibinfo {author} {\bibfnamefont {S.}~\bibnamefont {Pan}}, \bibinfo {author} {\bibfnamefont {X.}~\bibnamefont {Hu}}, \bibinfo {author} {\bibfnamefont {B.}~\bibnamefont {He}}, \ and\ \bibinfo {author} {\bibfnamefont {M.}~\bibnamefont {Huang}},\ }\href {\doibase 10.1016/J.ACTAMAT.2023.118993} {\bibfield  {journal} {\bibinfo  {journal} {Acta Materialia}\ }\textbf {\bibinfo {volume} {254}},\ \bibinfo {pages} {118993} (\bibinfo {year} {2023})}\BibitemShut {NoStop}%
\bibitem [{\citenamefont {Mazo}\ and\ \citenamefont {Kirkwood}(1955)}]{Mazo1955}%
  \BibitemOpen
  \bibfield  {author} {\bibinfo {author} {\bibfnamefont {R.~M.}\ \bibnamefont {Mazo}}\ and\ \bibinfo {author} {\bibfnamefont {J.~G.}\ \bibnamefont {Kirkwood}},\ }\href {\doibase 10.1073/PNAS.41.4.204} {\bibfield  {journal} {\bibinfo  {journal} {Proceedings of the National Academy of Sciences}\ }\textbf {\bibinfo {volume} {41}},\ \bibinfo {pages} {204} (\bibinfo {year} {1955})}\BibitemShut {NoStop}%
\bibitem [{\citenamefont {Koci}\ \emph {et~al.}(2007)\citenamefont {Koci}, \citenamefont {Ahuja}, \citenamefont {Belonoshko},\ and\ \citenamefont {Johansson}}]{Koci2007}%
  \BibitemOpen
  \bibfield  {author} {\bibinfo {author} {\bibfnamefont {L.}~\bibnamefont {Koci}}, \bibinfo {author} {\bibfnamefont {R.}~\bibnamefont {Ahuja}}, \bibinfo {author} {\bibfnamefont {A.~B.}\ \bibnamefont {Belonoshko}}, \ and\ \bibinfo {author} {\bibfnamefont {B.}~\bibnamefont {Johansson}},\ }\href {\doibase 10.1088/0953-8984/19/1/016206} {\bibfield  {journal} {\bibinfo  {journal} {Journal of Physics: Condensed Matter}\ }\textbf {\bibinfo {volume} {19}},\ \bibinfo {pages} {016206} (\bibinfo {year} {2007})}\BibitemShut {NoStop}%
\bibitem [{\citenamefont {Haghighat}, \citenamefont {Lucas},\ and\ \citenamefont {Schäublin}(2009)}]{Hafez2009}%
  \BibitemOpen
  \bibfield  {author} {\bibinfo {author} {\bibfnamefont {S.~M.~H.}\ \bibnamefont {Haghighat}}, \bibinfo {author} {\bibfnamefont {G.}~\bibnamefont {Lucas}}, \ and\ \bibinfo {author} {\bibfnamefont {R.}~\bibnamefont {Schäublin}},\ }\href {\doibase 10.1209/0295-5075/85/60008} {\bibfield  {journal} {\bibinfo  {journal} {EPL (Europhysics Letters)}\ }\textbf {\bibinfo {volume} {85}},\ \bibinfo {pages} {60008} (\bibinfo {year} {2009})}\BibitemShut {NoStop}%
\bibitem [{\citenamefont {Sharma}\ and\ \citenamefont {Parashar}(2024)}]{Sharma2024}%
  \BibitemOpen
  \bibfield  {author} {\bibinfo {author} {\bibfnamefont {S.~S.}\ \bibnamefont {Sharma}}\ and\ \bibinfo {author} {\bibfnamefont {A.}~\bibnamefont {Parashar}},\ }\href {\doibase 10.1088/1402-4896/AD7200} {\bibfield  {journal} {\bibinfo  {journal} {Physica Scripta}\ }\textbf {\bibinfo {volume} {99}},\ \bibinfo {pages} {102001} (\bibinfo {year} {2024})}\BibitemShut {NoStop}%
\bibitem [{\citenamefont {Donnelly}(1985)}]{Donnelly1985}%
  \BibitemOpen
  \bibfield  {author} {\bibinfo {author} {\bibfnamefont {S.~E.}\ \bibnamefont {Donnelly}},\ }\href {\doibase 10.1080/00337578508222514} {\bibfield  {journal} {\bibinfo  {journal} {Radiation Effects}\ }\textbf {\bibinfo {volume} {90}},\ \bibinfo {pages} {1} (\bibinfo {year} {1985})}\BibitemShut {NoStop}%
\bibitem [{\citenamefont {Schierholz}\ \emph {et~al.}(2015)\citenamefont {Schierholz}, \citenamefont {Lacroix}, \citenamefont {Godinho}, \citenamefont {Caballero-Hernández}, \citenamefont {Duchamp},\ and\ \citenamefont {Fernández}}]{Schierholz2015}%
  \BibitemOpen
  \bibfield  {author} {\bibinfo {author} {\bibfnamefont {R.}~\bibnamefont {Schierholz}}, \bibinfo {author} {\bibfnamefont {B.}~\bibnamefont {Lacroix}}, \bibinfo {author} {\bibfnamefont {V.}~\bibnamefont {Godinho}}, \bibinfo {author} {\bibfnamefont {J.}~\bibnamefont {Caballero-Hernández}}, \bibinfo {author} {\bibfnamefont {M.}~\bibnamefont {Duchamp}}, \ and\ \bibinfo {author} {\bibfnamefont {A.}~\bibnamefont {Fernández}},\ }\href {\doibase 10.1088/0957-4484/26/7/075703} {\bibfield  {journal} {\bibinfo  {journal} {Nanotechnology}\ }\textbf {\bibinfo {volume} {26}},\ \bibinfo {pages} {075703} (\bibinfo {year} {2015})}\BibitemShut {NoStop}%
\bibitem [{\citenamefont {Vitos}\ \emph {et~al.}(1998)\citenamefont {Vitos}, \citenamefont {Ruban}, \citenamefont {Skriver},\ and\ \citenamefont {Kollár}}]{Vitos1998}%
  \BibitemOpen
  \bibfield  {author} {\bibinfo {author} {\bibfnamefont {L.}~\bibnamefont {Vitos}}, \bibinfo {author} {\bibfnamefont {A.~V.}\ \bibnamefont {Ruban}}, \bibinfo {author} {\bibfnamefont {H.~L.}\ \bibnamefont {Skriver}}, \ and\ \bibinfo {author} {\bibfnamefont {J.}~\bibnamefont {Kollár}},\ }\href {\doibase 10.1016/S0039-6028(98)00363-X} {\bibfield  {journal} {\bibinfo  {journal} {Surface Science}\ }\textbf {\bibinfo {volume} {411}},\ \bibinfo {pages} {186} (\bibinfo {year} {1998})}\BibitemShut {NoStop}%
\bibitem [{\citenamefont {Farahani}\ and\ \citenamefont {Depla}(2024)}]{Farahani2024}%
  \BibitemOpen
  \bibfield  {author} {\bibinfo {author} {\bibfnamefont {F.~A.}\ \bibnamefont {Farahani}}\ and\ \bibinfo {author} {\bibfnamefont {D.}~\bibnamefont {Depla}},\ }\href {\doibase 10.1016/J.SURFCOAT.2024.131447} {\bibfield  {journal} {\bibinfo  {journal} {Surface and Coatings Technology}\ }\textbf {\bibinfo {volume} {494}},\ \bibinfo {pages} {131447} (\bibinfo {year} {2024})}\BibitemShut {NoStop}%
\bibitem [{\citenamefont {Farahani}\ \emph {et~al.}(2026)\citenamefont {Farahani}, \citenamefont {Akagawa}, \citenamefont {Shimizu},\ and\ \citenamefont {Depla}}]{Farahani2026}%
  \BibitemOpen
  \bibfield  {author} {\bibinfo {author} {\bibfnamefont {F.~A.}\ \bibnamefont {Farahani}}, \bibinfo {author} {\bibfnamefont {M.}~\bibnamefont {Akagawa}}, \bibinfo {author} {\bibfnamefont {T.}~\bibnamefont {Shimizu}}, \ and\ \bibinfo {author} {\bibfnamefont {D.}~\bibnamefont {Depla}},\ }\href {\doibase 10.1016/J.APSUSC.2025.164938} {\bibfield  {journal} {\bibinfo  {journal} {Applied Surface Science}\ }\textbf {\bibinfo {volume} {718}},\ \bibinfo {pages} {164938} (\bibinfo {year} {2026})}\BibitemShut {NoStop}%
\end{thebibliography}%

\end{document}